\documentclass[letterpaper,floatfix,aps,prb,amsmath,amssymb,twocolumn,superscriptaddress,nopacs,10pt]{revtex4-2}
\usepackage{graphicx}
\usepackage{verbatim}
\usepackage{xcolor}
\usepackage{array}
\usepackage{lipsum}
\usepackage{booktabs}
\usepackage{amsmath}
\usepackage{soul}
\usepackage[pdfauthor={La Magna, Catelani},
            pdftitle={Phonon down-conversion by normal metals for superconducting devices},
            breaklinks=true,
            colorlinks=true,
            linkcolor=blue,
            citecolor=blue,
            urlcolor=blue]{hyperref}

\hypersetup{pdfstartview={XYZ null null 1.05}}

\begin{document}
\title{Phonon down-conversion by normal metals for superconducting devices}

\author{G. La Magna}

\affiliation{Institute for Theoretical Nanoelectronics (PGI-2), Forschungszentrum J\"ulich, 52428 J\"ulich, Germany}

\affiliation{JARA Institute for Quantum Information, RWTH Aachen University, 52056 Aachen, Germany}

\author{G. Catelani}

\affiliation{Institute for Theoretical Nanoelectronics (PGI-2), Forschungszentrum J\"ulich, 52428 J\"ulich, Germany}

\affiliation{Quantum Research Center, Technology Innovation Institute, Abu Dhabi 9639, UAE}

\date{\today}

\begin{abstract}
Thanks to low dissipation, superconducting devices are promising for a number of applications, such as detectors and implementations of quantum computation. However, their working can be adversely impacted by quasiparticles, which is why so-called quasiparticle poisoning mechanisms and their mitigation are under intense investigation. Here we focus on one poisoning mechanism, namely pair-breaking phonons, and its mitigation through down-conversion by a normal-metal film -- the process in which scattering of high-energy phonons by electrons lowers the energy of the former below the pair-breaking threshold. To study the down-conversion, we introduce a model based on kinetic equations, which we solve both analytically (approximately) and numerically in the steady state. We use the solution {to} the estimate a properly-defined down-conversion efficiency which depends on material parameters (such as the strength of electron-phonon interaction and the phonon transmission coefficient at interfaces) and film and substrate thicknesses. Interestingly, we find that the efficiency is nearly optimal {above a minimum metal thickness,}
typically {between 50 and 500 nm depending on the metal-substrate combination.}

\end{abstract}

\maketitle

\section{Introduction}

Excess quasiparticles have long been recognized as a potential roadblock towards the implementation of superconducting quantum computers, as they can limit the coherence time of qubits~\cite{Aumentado2004,Catelani2011}. More recently, the focus has shifted on the role quasiparticles play in correlated error bursts that threatens the reliability of the quantum error correction schemes needed to achieve fault tolerance~\cite{Fowler_etal}.
In superconducting qubit arrays, the assumption that underpins error correction, namely that errors are uncorrelated, fails because of high‑energy impacts in the substrate from e.g. cosmic rays or environmental radioactivity;
 such an impact generates a cascade of electron–hole pairs which rapidly recombine, ultimately resulting in a burst of high‑energy phonons which propagate throughout the entire chip volume, reaching multiple qubits in a short time window.
These phonons then break Cooper pairs, producing athermal quasiparticles that cause error bursts~\cite{wilen_etal,McEwen_etal,Harrington2025}.

Numerous strategies are under consideration to mitigate the adverse effects of high-energy impacts. For instance, operating in a deep underground facility~\cite{cardani_etal,Cardani2023,DeDomincis2026} or shielding the cryostat containing the chip with thick lead layers~\cite{vepsalainen_etal} decreases the rate of impacts; however, these solutions can be impractical and a more desirable approach would be to design resilient chips. This goal can be achieved in several ways: normal metals~\cite{Riwar2016} or low-gap superconductors~\cite{Riwar2019,Pan2022} in contact with the qubit can act as quasiparticle traps; gap engineering at the qubit junctions can hinder quasiparticle tunneling~\cite{Marchegiani2022} and consequently suppress relaxation errors~\cite{McEwen2024}, although residual phase errors are still present~\cite{Kurilovich2026}. Low-gap superconducting islands~\cite{henriques.2019} or normal-metal ones~\cite{iaia_etal,Larson2025} can be used to convert high-energy pair-breaking phonons down to low energy -- this is the approach that we study in this work.

The process of phonon down-conversion consists of four basic steps: 1. a high-energy phonon is transmitted from the substrates into the metal; 2. in the metal, the phonon excites an electron-hole pair (or breaks a Cooper pair in a low-gap superconductor); 3. the electronic excitations relax by emitting phonons of lower energy; 4. the low-energy phonons are transmitted back into the substrate. A recent work~\cite{Modeling} simulated this chain of events using a Monte-Carlo approach -- in fact, the simulation can start even from earlier steps, such as the generation of electron-hole pairs by gamma ray impacts. This numerical technique can take into account the actual geometry of the device (for instance, lateral dimensions of the chip and location of qubits on top of the substrate). Here we take a different approach, based on kinetic equations for the distribution functions of phonons and electrons; while in principle spatial dependencies can be taken into account, we ignore them for simplicity. The model takes into consideration phonon transmission at the substrate-metal interface (steps 1 and 4) and the electron-phonon interaction responsible for down-conversion (steps 2 and 3). An advantage of our analytical methodology is that it readily enables {estimating the} down-conversion efficiency -- a quantity defined in terms of quasiparticle generation rates in the qubit in the presence and absence of normal metal --  {as function of metal thickness and for} 
different combinations of materials for the substrate and the metal, thus aiding in {chip design decisions} 
for the purpose of mitigating quasiparticle poisoning.

The paper is organized as follows: in the next section we introduce the kinetic equation-based model for the dynamics of electron and phonons and the relevant physical parameters, whose estimation is the subject of Sec.~\ref{sec:par_est}. The steady-state solution of the kinetic equations is considered in Sec.~\ref{sec:distributions}, while Sec.~\ref{sect:DW_eff} studies the down-conversion efficiency. Our  conclusions are presented in Sec.~\ref{sec:conclusions}, which is followed by Appendices~\ref{app:eeint} through \ref{app:Rqp_minimum}.

\section{Model} 
\label{sec:model}

\begin{figure}[!tb]
    \centering
\includegraphics[width=1.0\columnwidth]{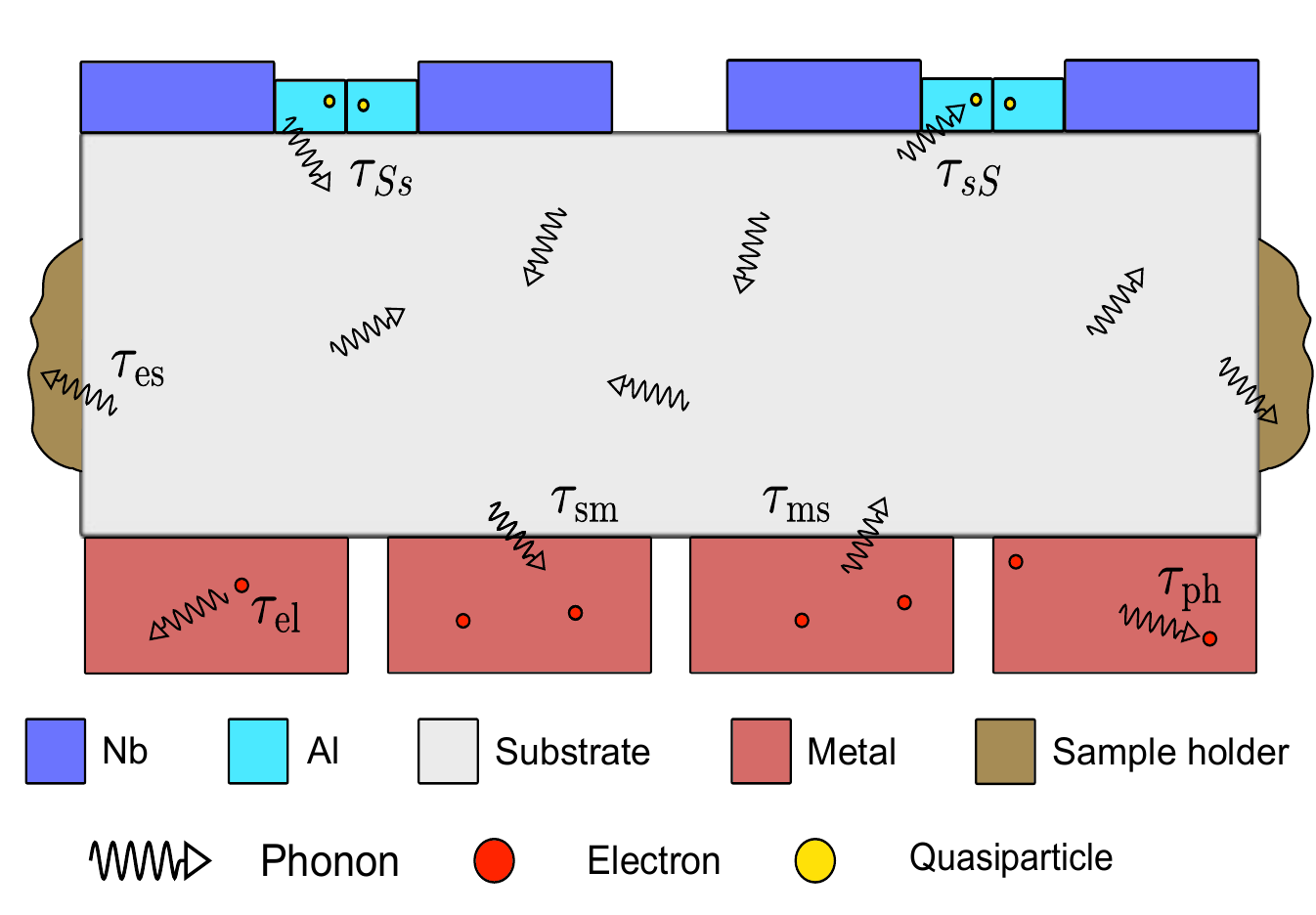}
    \caption{
    Schematic depiction of a superconducting qubit chip with normal metal islands on the backside. The physical processes included in the kinetic equations are all represented: phonon escape to the sample holder (lifetime $\tau_\mathrm{es}$); phonon transmission from substrate to metal ($\tau_{sm}$) and from metal to substrate ($\tau_{ms}$); electron-phonon interaction ($\tau_\mathrm{el}$ and $\tau_\mathrm{ph}$). 
    Phonon transmission at the substrate-superconductor interface ($\tau_{sS}$ and $\tau_{Ss}$) has to be accounted for in the calculation of the down-conversion efficiency.}
    \label{fig:model}
\end{figure}

{In a typical device comprising superconducting qubits, the superconducting circuits are fabricated on top of a dielectric substrate, typically silicon, which constitutes the bulk of the chip volume. For down-converting phonons, a normal-metal film is deposited on the back of the chip, and the chip is thermally coupled to the cryostat through a sample holder.}
To 
study the down-conversion efficiency by a normal-metal film, we consider a simplified model of a chip consisting of 
a dielectric substrate of thickness $d_s$ with a metallic film of thickness $d_m$ deposited on its back side. 
For simplicity, we do not include the superconducting qubits and the ground plane on the front side in the model. The first simplification is in general justified, since typically the relevant qubit structures have much smaller area and volume than the metal on the back side. The second simplification applies in the case of ground plane with gap larger than that of the qubit junctions' superconductor (e.g. Nb ground plane and Al junctions) and phonons with energy between the two gaps; the model is easily generalizable to include the effect of the ground plane if necessary.
Finally, we take into account the possibility that
phonons escape from the chip through its connections to a sample holder, such as wire bonds and/or glues.
A schematic of the modeled device is shown in Fig.~\ref{fig:model}

We adopt a kinetic equation approach to model 
phonon down-conversion in the metallic film.
Here we assume that phonons are 
generated in
the substrate at a rate
$g(\Omega,t)$, which in general is a function of both phonon {energy} $\Omega$ and time $t$. This rate can account for phonons 
resulting from the rapid recombination of electron–hole pairs randomly generated at radiation impact sites, followed by
propagation through the substrate, and/or for phonons intentionally generated
in a controlled manner by properly biasing  SIS~\cite{iaia_etal} or NIS~\cite{patel.2017} junctions.

To simplify the analysis, we average the phonon distributions over polarization, space, and momentum direction. We perform the analogous spatial and momentum averages for the electron distribution, and we assume particle--hole symmetry for $f(E,t)$. With these assumptions, spatial gradients are neglected and the system is described by coupled kinetic equations for the substrate phonons $n_s(\Omega,t)$, the metal phonons $n(\Omega,t)$, and the electrons $f(E,t)$:
\begin{align}\label{eq:sist_eq}
  \frac{\partial n_{s}}{\partial t}
    &= g-\frac{n_{s}-n_B}{\tau_\mathrm{es}}-{\frac{n_{s}-n}{\tau_{sm}}}, \\
  \frac{\partial n}{\partial t}
    &= I_{\rm ph}\{n,f\}-{\frac{n-n_s}{\tau_{ms}}}, \\
  \frac{\partial f}{\partial t}
    &= I_{\rm el}\{n,f\}, \label{eq:keq_f}
\end{align}
where $n_B(\Omega,T)={1/[e^{\Omega/(k_B T)}-1]}$ is the equilibrium phonon distribution of the sample holder (bath) at temperature $T$, $\tau_\mathrm{es}^{-1}$ is the effective escape rate of substrate phonons to the sample holder, and $\tau_{sm}^{-1}$ ($\tau_{ms}^{-1}$) are the substrate to metal (metal to substrate) transmission rates; {the relation between the two rates is discussed in Sec.~\ref{sec:mstr}}. 
The collision integrals for electron-phonon interactions can be derived from Fermi’s golden rule (see for instance Ref.~\cite{Electron_rel})
\begin{widetext}
\begin{align}
I_{\mathrm{el}}\{n,f\}
= & \frac{1}{\tau_\mathrm{el}\,\Omega_{D}^{3}}
\biggl\{
  \int_{0}^{\Omega_{D}}\!\!d\Omega\,\Omega^{2}\left[f(E{+}\Omega)\,(1{-}f(E))(1{+}n(\Omega)) 
- f(E)\,(1{-}f(E{+}\Omega))\,n(\Omega)\right]\notag\\
&\qquad \quad
  +\int_{0}^{E}\!d\Omega\,\Omega^{2}\left[f(E{-}\Omega)\,(1{-}f(E))\,n(\Omega)
  -f(E)\,(1{-}f(E{-}\Omega))\,(1{+}n(\Omega))\right]
\notag\\ &\qquad \quad
+\int_{E}^{\Omega_{D}}\!\!\!d\Omega\,\Omega^{2}\left[(1{-}f(\Omega{-}E))\,(1{-}f(E))\,n(\Omega)
-f(E)\,f(\Omega{-}E)\,(1{+}n(\Omega))\right]
\biggr\}, \label{eq:Iel}
\end{align}
\begin{align}
I_{\mathrm{ph}}\{n,f\}
= & \frac{1}{\tau_{\mathrm{ph}}\,\Omega_{D}}
\biggl\{
  2\!\int_{0}^{\Omega_{D}}\!dE\,
    \left[f(E{+}\Omega)\,(1{-}f(E))(1{+}n(\Omega))  
- f(E)\,(1{-}f(E{+}\Omega))\,n(\Omega)\right]\notag\\
&\qquad\quad\
  +\int_{0}^{\Omega}\!dE\,
    \left[f(E)\,f(\Omega{-}E)(1{+}n(\Omega)) 
-(1{-}f(E))\,(1{-}f(\Omega{-}E))\,n(\Omega)\right]
\biggr\}, \label{eq:Iph}
\end{align}
\end{widetext}
where $\tau_\mathrm{el}$ is (one third of) the lifetime of an electron at the Debye energy $\Omega_D$ with respect {to} phonon emission, while $\tau_\mathrm{ph}$  denotes the lifetime of a phonon at the Debye energy with respect the creation of an electron-hole pair; the two lifetimes are related, as we detail in Sec.~\ref{sec:E-PH}. The first two lines in Eq.~\eqref{eq:Iel} account for emission and absorption of phonons by the electrons; the {first} term in the last line describes phonon absorption with creation of an electron-hole pair and the {last} term in that line the inverse process. The two lines in Eq.~\eqref{eq:Iph} describe the same processes for the phonons. 
Note that by adopting the Debye model for phonons (see also Sec.~\ref{sec:E-PH}), we ignore anharmonic effects such as phonon-phonon scattering, which are strongly suppressed at the low energies we are interested in~\cite{Tamura}. We additionally neglect electron-electron scattering processes in the kinetic equation for $f$; we comment on the applicability of this approximation in Appendix~\ref{app:eeint}.

\subsection{Steady state}

In this work we consider
the steady-state solution of the kinetic equations, 
assuming $T=0$ for the bath and a time-independent phonon generation rate $g$:
\begin{align}
    g(\Omega) - \frac{n_s(\Omega)}{\tau_\mathrm{es}} - {\frac{n_s(\Omega)-n(\Omega)}{\tau_{sm}}} &= 0, \label{eq:ns_st} \\
    I_{\text{ph}}\{n, f\}  - {\frac{n(\Omega)-n_s(\Omega)}{\tau_{ms}}} &= 0, \label{eq:n_st} \\
    I_{\text{el}}\{n, f\} &= 0. \label{eq:f_st}
\end{align}
The first equation in this system can be readily solved to express the phonon distribution $n_s$ in the substrate 
in terms of the phonon distribution $n$ in the metallic film,
\begin{equation}\label{eq:phonon_substrate}
n_s(\Omega) = \frac{\tau_{sm}\tau_\mathrm{es}}{\tau_{sm}+\tau_\mathrm{es}} \left[ g(\Omega) + \frac{n(\Omega)}{{\tau_{sm}}} \right].    
\end{equation}

In the \emph{no-metal} limit ($\tau_{sm} \to \infty$), the steady-state phonon distribution in the substrate reduces to
\begin{equation}\label{eq:phonon_substrate_no_metal}
    n_{s}^\mathrm{nm}(\Omega) = \tau_\mathrm{es}g(\Omega).
\end{equation}
In the absence of the metal, the phonon distribution is determined solely by the balance between phonon generation and their escape from the substrate.  
Conversely, in the limit of \emph{perfect absorption} by the metal {we set $n(\Omega)=0$ in Eq.~\eqref{eq:ns_st} and} 
the steady-state distribution in the substrate is given by
\begin{equation}\label{eq:perf_absorp}
    n_{s}^\mathrm{pa}(\Omega)  = \frac{\tau_{sm}\tau_\mathrm{es}}{\tau_{sm}+\tau_\mathrm{es}} g(\Omega).
\end{equation}
In this case the metal layer acts as an additional zero-temperature bath;
this results in a reduction of the phonon population relative to the case of a substrate alone, as quantified by the factor $ \frac{\tau_{sm}}{\tau_{sm}+\tau_\mathrm{es}} < 1$. 
These two limiting expressions will be useful when considering the down-conversion efficiency in Sec.~\ref{sect:DW_eff}.

In the general case in which both $\tau_{sm}$ and {$n(\Omega)$ in Eq.~\eqref{eq:ns_st}} are finite,
phonons can be down-converted within the metal and subsequently transmitted back into the substrate,
as accounted for by the second term in square brackets
in Eq.~\eqref{eq:phonon_substrate}.
Therefore, knowing the distribution $n(\Omega)$ will be crucial in determining the down-conversion efficiency of the metal layer.
Substituting the expression for the substrate phonon distribution, Eq.~\eqref{eq:phonon_substrate}, into Eq.~\eqref{eq:n_st} yields:
\begin{equation}\label{eq:k_eq_eff}
I_{\text{ph}}\{n, f\} - \frac{n(\Omega) - n_{\text{inj}}(\Omega)}{\overline{\tau}_{ms}} = 0 \, ,
\end{equation}
where we have introduced the notation
\begin{align}\label{eq:effective_parameters}
\frac{1}{\overline{\tau}_{ms}} & = \frac{1}{\tau_{ms}} \left( 1 - \frac{\tau_\mathrm{es}}{\tau_{sm}+\tau_\mathrm{es}} \right), \\
 n_{\text{inj}}(\Omega) & = 
 {\tau_\mathrm{es}g(\Omega)}, \label{eq:ninj}
\end{align}
for the effective escape rate $\overline\tau_{ms}^{-1}$ of phonons from the metal (the factor in brackets account for the possibility that phonons return to the metal from the substrate before they escape to the sample holder) and the phonon population $n_\mathrm{inj}$ injected into the metal. Our goal is to solve Eq.~\eqref{eq:k_eq_eff} together with Eq.~\eqref{eq:f_st} to then obtain $n_s$ from Eq.~\eqref{eq:phonon_substrate}. For practical applications, we need to consider realistic values of the parameters entering these equations, whose estimates are the subject of the next section.

\section{Parameter estimation}
\label{sec:par_est}

In this section we discuss estimates for the timescales entering into our model, namely the lifetimes for transmission from metal to substrate and from substrate to metal, $\tau_{ms}$ and $\tau_{sm}$; the substrate to sample holder escape time, $\tau_\mathrm{es}$; and the times characterizing the electron-phonon interaction strength, $\tau_\mathrm{el}$ and $\tau_\mathrm{ph}$. As these times depend on material properties, for concreteness we consider two typical substrate materials, Si and $\text{Al}_2\text{O}_3$, and a few metals: Al, Nb, Ti, Cu, Au, Ag, and Ta. Although we consider here down-conversion in a normal metal, we include some superconducting materials for two reasons: first, at temperatures between the two critical ones of two metals, the lower-$T_c$ film is in the normal state, so our model applies to that case; second, these parameters are needed if one wants to model down-conversion by a superconducting film. We consider first the phonon transmission between metal and substrate.

\subsection{Metal-substrate transmission}\label{sec:mstr}

The phonon transmission lifetimes from the substrate to the metal, $\tau_{sm}$, and from the metal to the substrate, $\tau_{ms}$, depend on material properties and geometry as follows~\cite{kaplan}:
\begin{equation}\label{eq:transmission_times}
\tau_{sm} = \frac{4 d_{s}}{c_{s} \eta_{sm}}, \quad
\tau_{ms} = \frac{4 d_{m}}{c_{m} \eta_{ms}}, 
\end{equation}
where $ d_{\alpha}$ denotes the thickness of the metal ($\alpha=m$) or the substrate ($\alpha=s$), $c_\alpha$ the speed of sound, and $\eta_{\alpha\beta}$ the transmission coefficient from $\alpha$ to $\beta$. The last two quantities are averages (over polarization and for transmission over momentum direction), as we will explain in what follows.

\begin{table*}[!tbh]
\centering
\begin{tabular}{@{}lc|rrrrrrrrr@{}}
\toprule
&  & Al $ $ & Nb $ $ & Ti $ $ & Cu $ $ & Au $ $ & Ag $ $ & Ta $ $ & Al$_2$O$_3$ & Si $ $ \\
\midrule
$\rho$  & [g/cm$^3$]       & $\,$2.733 & $\,$8.593 & $\,$4.522 & $\,$9.030 & 19.300 & 10.630 & 16.173 & 3.992 & $\,$2.332 \\
$c_L$ & [mm/$\mu$s]       & 6.808 & 5.139 & 6.253 & 4.828 & 3.377  & 3.788  & 4.179 & 10.894 & 8.997 \\
$c_T$ & [mm/$\mu$s]      & 3.250 & 2.168 & 3.330 & 2.380 & 1.242  & 1.756  & 2.093  & 6.453 & 5.366 \\
$c$  & [mm/$\mu$s]         & {3.434} & 2.276 & 3.535 & 2.518 & 1.294  & 1.852  & 2.216 & 6.871 & 5.714 \\
$F_T$      &             & 0.948 & 0.964 & 0.930 & 0.943 & 0.976  & 0.953  & 0.941 & 0.906 & 0.904\\
$F_L$       &            & 0.052 & 0.036 & 0.070 & 0.057 & 0.024  & 0.047  & 0.059  & 0.094 & 0.096 \\
\bottomrule
\end{tabular}
\caption{Properties of selected materials. Values in the first three rows are taken from Ref.~\cite{Simmons_} and used to calculate those in the last three rows as explained in the text.}
\label{tab:material_properties}
\end{table*}

For the speed of sound $c$, we weight the contributions from the two transverse modes and the longitudinal one by the respective fractional density of states  $F_T$ and $F_L$ (for notational simplicity, we drop in $c_\alpha$, $c_{T\alpha}$ and $c_{L\alpha}$ the subscript $\alpha$ denoting a particular material),
\begin{equation}
    c = F_{T} c_{T} + F_{L} c_{L} .
\end{equation}
Within the Debye model, which we employ in this work, they can be expressed in terms of the transverse and longitudinal sound speeds $c_T$ and $c_L$ as $F_T = 2c_L^3/(c_T^3+2c_L^3)$ and $F_L=1-F_T$. For reference, we report in Table~\ref{tab:material_properties} the values of longitudinal and transverse sound speeds (as well as density, needed to calculate transmission coefficients) from which we obtain the average phonon velocity and the fractional densities of states.

The transmission coefficient is similarly expressed as a weighted average,
\begin{equation}\label{eq:etaav}
\eta = {\frac{F_L c_L \eta_L + F_T c_T \eta_T}{c}}    
\end{equation}
Two models are widely used to calculate the transmission coefficient, namely the  Acoustic Mismatch Model (AMM) and the Diffuse Mismatch Model (DMM)~\cite{Thermal,Interfacial}. Their validity has been investigated largely through measurements of thermal interfacial resistance; broadly, the DMM applies at higher temperature and hence to higher-energy phonons and to rougher interfaces in comparison to the AMM, which should {be} therefore preferred in the low-temperature setting of superconducting devices. However, the calculation of $\eta$ in the DMM is much simpler, and the results of the two models are comparable, so we report estimates using both models.

In the DMM, phonons scattering at an interface are assumed to lose memory of their direction and as a consequence the relation $\eta_{\alpha\beta}+\eta_{\beta\alpha}=1$ holds. Then 
{requiring detailed balance to hold} one finds
\begin{equation}\label{eq:etaDMM}
\eta_{\alpha\beta} = \frac{\sum_{\mu} c_{\mu\beta}^{-2}}{\sum_{\mu} \left( c_{\mu\beta}^{-2} + c_{\mu\alpha}^{-2} \right) },    
\end{equation}
where the sums over index $\mu$ account for the longitudinal ($\mu=L$) and two transverse ($\mu=T_1,T_2$) polarizations.

In the AMM, phonons are treated as elastic waves propagating through an isotropic, continuous medium, and the interface between the two solids is idealized as a perfectly smooth and planar surface. An incident phonon can then reflect and refract, as well as mode-convert (between longitudinal and transverse polarizations). Suitable continuity conditions for displacement and stress tensor at the interface result in the equivalent of Snell's law of refraction, with the inverse speed of sound playing the role of the refraction index, and hence in the existence of critical angles above which the incident phonon is not transmitted if the speed of sound in the medium of the incident phonon is smaller than that in the second medium. Transverse waves with displacement parallel to the surface {($SH$ waves)} do not mode-convert, and their transmission coefficient from material $\alpha$ to material $\beta$ is
\begin{equation}
\eta_{T\alpha\beta}^{{SH}}(\theta_i) = \frac{4Z_\alpha Z_\beta \cos{\theta_i} \cos{\theta_r}}{(Z_\alpha \cos{\theta_i} + Z_\beta \cos{\theta_r})^2},    
\end{equation}
where $Z = \rho c_{T}$, with $\rho$ the density, is the acoustic impedance, $\theta_i$ the incidence angle, and $\theta_r$ the refraction one. This angle-dependent transmission coefficient is then averaged over all possible incidence angles (possibly up to the critical one). Longitudinal and transverse waves with displacement not parallel to the interface {($SV$ waves)} can mode-convert, so the calculation of their transmission coefficient is more involved, but it can be implemented numerically~\cite{kaplan}. We give in Table~\ref{tab:amm_coefficients} our results for the longitudinal and transverse {[$\eta_T = (\eta_T^{SH}+\eta_T^{SV})/2$~\cite{kaplan}]} transmission coefficients between a few metals and two substrate materials~\footnote{{The transmission coefficients satisfy the reciprocity relation $\eta_{L\alpha\beta}/c^2_{L\alpha}+2\eta_{T\alpha\beta}/c^2_{T\alpha} = (\alpha\leftrightarrow \beta)$~\cite{kaplan}; together with the definition in Eq.~\eqref{eq:etaav}, this implies that detailed balance holds for the AMM as well.}}. For three metals (Al, Nb, Ta) to substrate transmission, our results are comparable to those reported in Ref.~\cite{kaplan}; the slight differences are attributable to the used values for density and sound speeds. In Table~\ref{tab:avg_coefficients} we collect for both models the estimates for the average transmission coefficients which can be obtained from the data in the first two tables.

\begin{table}[tb]
\centering
\resizebox{\columnwidth}{!}{\begin{tabular}{l|cc|cc}
\toprule
& Al$_2$O$_3$ & Si & Al$_2$O$_3$ & Si \\
\midrule
Al & {(0.509, 0.192)}
   & {(0.647, 0.331)}
   & {(0.852, 0.837)}
   & {(0.980, 0.928)} \\
Nb & {(0.379, 0.091)}
   & {(0.425, 0.119)}
   & {(0.974, 0.933)}
   & {(0.835, 0.813)} \\
Ti & {(0.487, 0.223)}
   & {(0.568, 0.333)}
   & {(0.950, 0.930)}
   & {(0.939, 0.906)} \\
Cu & {(0.336, 0.109)}
   & {(0.387, 0.137)}
   & {(0.970, 0.930)}
   & {(0.826, 0.790)} \\
Au & {(0.169, 0.028)}
   & {(0.175, 0.033)}
   & {(0.931, 0.903)}
   & {(0.710, 0.716)} \\
Ag & {(0.226, 0.058)}
   & {(0.265, 0.075)}
   & {(0.973, 0.935)}
   & {(0.853, 0.817)} \\
Ta & {(0.240, 0.077)}
   & {(0.257, 0.082)}
   & {(0.901, 0.856)}
   & {(0.667, 0.635)} \\
\bottomrule
\end{tabular}}
\caption{Transmission coefficients
$(\eta_{Lms},\eta_{Tms})$ (left two columns) and
$(\eta_{Lsm},\eta_{Tsm})$ (right two columns) in the AMM for different combinations of metals and dielectric substrates.}
\label{tab:amm_coefficients}
\end{table}

\begin{table}[tb]
\centering
\resizebox{\columnwidth}{!}{\begin{tabular}{l|cc|cc}
\toprule
& Al$_2$O$_3$ & Si & Al$_2$O$_3$ & Si \\
\midrule
Al & {(0.225, 0.840)}
   & {(0.363, 0.936)}
   & {(0.211, 0.789)} & {(0.279, 0.721)} \\
Nb & {(0.114, 0.939)}
   & {(0.144, 0.816)}
   & {(0.109, 0.891)} & {(0.150, 0.850)} \\
Ti & {(0.256, 0.933)}
   & {(0.362, 0.911)}
   & {(0.215, 0.785)} & {(0.284, 0.716)} \\
Cu & {(0.133, 0.936)}
   & {(0.164, 0.795)}
   & {(0.125, 0.875)} & {(0.171, 0.829)} \\
Au & {(0.037, 0.907)}
   & {(0.042, 0.715)}
   & {(0.039, 0.961)} & {(0.056, 0.944)} \\
Ag & {(0.074, 0.941)}
   & {(0.094, 0.822)}
   & {(0.073, 0.927)} & {(0.102, 0.898)} \\
Ta & {(0.095, 0.863)}
   & {(0.102, 0.640)}
   & {(0.099, 0.901)} & {(0.137, 0.863)} \\
\bottomrule
\end{tabular}%
}
\caption{Transmission coefficients $(\eta_{ms}, \eta_{sm})$ in the AMM (left two columns) and the DMM (right two columns) for different combinations of metals and dielectric substrates.}
\label{tab:avg_coefficients}
\end{table}

Having obtained estimates for the (average) sound speeds and transmission coefficients, we can now evaluate the lifetimes for transmission of phonons from substrate to metal, $\tau_{sm}$, and from metal to substrate, $\tau_{ms}$, using Eq.~\eqref{eq:transmission_times}.
We take $d_{s} = 500 \, \mu\mathrm{m}$ for the substrate thickness and $d_{m} = 5 \, \mu\mathrm{m}$ for the metal thickness, which are typical values employed in experiments (see Ref.~\cite{Modeling} and references there); we will further discuss the choice of $d_m$ in Sec.~\ref{sect:DW_eff}. 
In actual devices the backside metallization is patterned into an array of separated metallic islands in order to suppress qubit relaxation caused by ohmic losses in the metal that is capacitively coupled to the qubit~\cite{martinis}. For instance, in Refs.~\cite{iaia_etal,Modeling}, the backside metallization consists of arrays of $200\,\mu\mathrm{m} \times 200\,\mu\mathrm{m}$ metallic islands separated by $50\,\mu\mathrm{m}$, corresponding to a coverage of about $64\%$ of the backside surface; the role of varying coverage fraction (from 0 to $34\%$) has been investigated in Ref.~\cite{henriques.2019}, although using superconducting islands on the top of the chip. 
To account for the coverage fraction $p$, we renormalize the transmission coefficient from substrate to metal in the expression for $\tau_{sm}$, Eq.~\eqref{eq:transmission_times}, as $\eta_{sm} \to p\,\eta_{sm}$;
here we set $p=0.64$ as a representative value.
The transmission lifetimes obtained from Eq.~\eqref{eq:transmission_times} with our choice of parameters for various combinations of metal and substrate are reported in Table~\ref{tab:tau_ell}. Note that the two models, AMM and DMM, lead to similar results for the lifetimes, with typical deviations
{of less than 10\% for Al$_2$O$_3$ and up to $\sim$30\% for Si}.

\begin{table}[tb]
\centering
\begin{tabular}{l|cc|cc}
\toprule
& Al$_2$O$_3$ & Si & Al$_2$O$_3$ & Si \\
\midrule
Al & \phantom{1}({26}, {542})
   & \phantom{1}({16}, {584})
   & \phantom{1}(28, 577)
   & \phantom{1}(21, 759) \\
Nb & \phantom{1}({77}, {484})
   & \phantom{1}({61}, {670})
   & \phantom{1}(81, 510)
   & \phantom{1}(59, 643) \\
Ti & \phantom{1}({22}, {487})
   & \phantom{1}({16}, {600})
   & \phantom{1}(26, 579)
   & \phantom{1}(20, 764) \\
Cu & \phantom{1}({60}, {486})
   & \phantom{1}({48}, {688})
   & \phantom{1}(64, 520)
   & \phantom{1}(46, 660) \\
Au & ({418}, {501})
   & ({366}, {765})
   & (394, 473)
   & (277, 579) \\
Ag & ({146}, {484})
   & ({115}, {665})
   & (148, 491)
   & (106, 609) \\
Ta & \phantom{1}({95}, {527})
   & \phantom{1}({89}, {855})
   & \phantom{1}(91, 505)
   & \phantom{1}(66, 634) \\
\bottomrule
\end{tabular}
\caption{Transmission lifetimes $({\tau_{ms}}, {\tau_{sm}})$ in ns computed using in Eq.~\eqref{eq:transmission_times} the transmission coefficients from the AMM (left two columns) and from the DMM (right two columns) of Table~\ref{tab:avg_coefficients} (up to the renormalization due to partial coverage discussed in the text), the speed of sound $c$ of Table~\ref{tab:material_properties}, and $d_s=500\,\mu$m and $d_m=5\,\mu$m for the substrate and metal thicknesses.}
\label{tab:tau_ell}
\end{table}

\subsection{Substrate phonon escape time}\label{sec:Phonon escape time}

Cryogenic devices are usually inserted into a sample holder which is then thermally coupled to the base plate of a refrigerator. We assume the sample holder to be at thermal equilibrium at a temperature much smaller than all other energy scales in the problem, so that it effectively acts as a sink for phonons. The escape rate from substrate to sample holder depends on how the device is connected to the latter; for example, the device can be suspended using wire-bonds, glued to the holder, or clamped. 

For suspended devices, the escape rate can be estimated as~\cite{martinis}
\begin{equation}
\frac{1}{\tau_\mathrm{es}} = \left(\frac{N_w A}{V}\right) c_s \left(\frac{l}{L}\right),
\end{equation}
where $N_w$ is the number of wire-bonds, $A$ their cross-sectional area, $L$ their length, $V$ the substrate volume, and $l$ the phonon mean free path in the wire-bonds (which in the reference was taken to be of order $l \sim \sqrt{A}$). For the typical parameters quoted in Ref.~\cite{martinis}, this expression gives  $\tau_\mathrm{es}$ of order few ms. We are not aware of experimental determination of $\tau_\mathrm{es}$ for such a setup, and since we show below that shorter $\tau_\mathrm{es}$ can be achieved, we will not discuss this approach further. 

We next consider chips glued to the sample holder. As typical examples we take Refs.~\cite{iaia_etal,Modeling}, in which the  corners of the chip were attached to the sample holder with GE varnish. By monitoring the change in qubit relaxation, the change in quasiparticle density can be estimated; as we discuss in more detail in Appendix~\ref{app:escape}, its decay at long time is dominated by the escape of phonons from the substrate, which enables us to extract $\tau_\mathrm{es}$ directly from the data; fitting an exponential decay at long times (which we define as times after which the excess quasiparticle density has decayed to below one quarter from its peak), we find $\tau_\mathrm{es}$ between 35 and 105$\,\mu$s. As a check to our approach, we note that based on simulations of phonon and quasiparticle dynamics, Ref.~\cite{Modeling} reported a phonon escape probability from the substrate of 2.5\%. Given this probability, the lateral dimension of the chip (8~mm), and the speed of sound in the Si substrate, a rough estimate of the escape time is in the tens of $\mu$s, compatible with our findings from directly fitting the data.

A similar fitting approach to that discussed above was recently used for clamped chips in Ref.~\cite{nho2025}, resulting in $\tau_\mathrm{es}\simeq0.7\,$ms. It was noted there that while the qubit relaxation rate recovers after a burst event over this time scale, the excitation rate and hence the qubit effective temperature return to their pre-burst values over a time roughly one order of magnitude longer. This could indicate that phonons with energy below the pair-breaking threshold leave the chip more slowly; we do not explore the implications of this slower timescale, as this goes beyond the scope of the present work.

\subsection{Electron--phonon interaction}\label{sec:E-PH}

The electron--phonon (e--ph) interaction enters the kinetic equations through the 
collision integrals $I_{\mathrm{el}}$ and $I_{\mathrm{ph}}$ in 
Eqs.~\eqref{eq:Iel} and \eqref{eq:Iph}. The kernel in Eq.~\eqref{eq:Iel} is $\Omega^2/(\tau_\mathrm{el}\Omega_D^3)$, but more generally it is given by {$2\pi/\hbar$ times} the Eliashberg function $\alpha^{2}(\Omega)F(\Omega)$, defined as 
the Fermi--surface average of the squared e--ph matrix element at phonon energy 
$\Omega$. 
At low energies and for sufficiently clean metals, $\alpha^{2}(\Omega)F(\Omega)$ is well 
approximated by a quadratic law~\cite{kaplan},
\begin{equation}
\alpha^{2}(\Omega)F(\Omega) \simeq b\,\Omega^{2},
\label{eq:A2F_quadratic}
\end{equation}
where, since the {Eliashberg function} is dimensionless, $b$ has units of inverse energy squared 
(we consider briefly the case of disordered metals in Appendix~\ref{app:dirty_eph}).
It is convenient to characterize the electron-phonon coupling by the timescale
\begin{equation}
\tau_{\rm el}
\equiv
\frac{\hbar}{2\pi b\Omega_D^{3}},
\label{eq:tau_el_def}
\end{equation}
where $\Omega_D$ is the Debye energy. 
For superconducting metals, this time is proportional to $\tau_0$ of Ref.~\cite{kaplan1976}, $\tau_\mathrm{el}= \tau_0 (k_B T_c/\Omega_D)^3/Z_1(0)$, with $T_c$ the critical temperature and $Z_1(0)$ the so-called renormalization parameter~\cite{kaplan1976}. 

To determine $b$ and hence $\tau_\mathrm{el}$, we start by considering
the mass-enhancement factor
\begin{equation}
\lambda_{\rm m} = 2\!\int_{0}^{\Omega_D} 
d\Omega\,\frac{\alpha^{2}(\Omega)F(\Omega)}{\Omega},
\label{eq:lambda_def}
\end{equation}
whose values, estimated both from theory and experiments, are available for most elemental metals~\cite{GGrimvall_1976}.  
Since the ratio $[\alpha^{2}(\Omega)F(\Omega)]/F(\Omega)$ remains approximately constant at low energies~\cite{GGrimvall_1976}, we define an effective coupling constant via
\begin{equation}
\overline{\alpha}^{2}
=
\frac{\lambda_{\rm m}}{2\!\displaystyle\int_0^{\Omega_D}
\frac{F(\Omega)}{\Omega}\,d\Omega}.
\label{eq:a2bar_def}
\end{equation}
To evaluate the denominator, we employ phonon densities of states from Born--von K\'arm\'an lattice-dynamics 
calculations~\cite{LandoltBoernstein}, 
normalized to $\int F(\Omega)\,d\Omega=3$~\cite{kaplan1976}. 
The results of these {calculations} are reported in the second row of Table~\ref{tab:eph_prefactors}.

At low energies, the phonon density of states (DoS), which has units of inverse energy, also follows a quadratic law,
\begin{equation}
F(\Omega) \simeq \beta\,\Omega^{2},
\label{eq:F_quadratic}
\end{equation}
where the coefficient $\beta$ has units of inverse energy to the third power and within the Debye model it is given by
\begin{equation}\label{eq:betaD}
   {\beta = \frac{c_L^{-3}+2c_T^{-3}}{2\pi^2 \hbar^3 N}}
\end{equation}
with $N$ the atomic number density; both $N$ and $\beta$ calculated from it are reported in Table~\ref{tab:eph_prefactors}. For comparison, we also estimate $\beta$ by fitting
the normalized phonon DoS of Ref.~\cite{LandoltBoernstein} to the quadratic law $\Omega_0 F(\Omega) = \tilde{\beta}(\Omega/\Omega_0)^{2}$ (with $\Omega_0 = 1\,$meV) in the energy window $0.5 \lesssim \Omega \lesssim 5$\,meV, well below features {that cannot be accounted for within the Debye model (such as non-linear phonon dispersion);}
we take the upper energy bound as limiting in practice the applicability of the Debye model to energies up to a fraction of the Debye energy;
we then identify $\beta_\mathrm{fit} = \tilde{\beta}/\Omega_0^3$. {The two estimates for $\beta$ deviate from each other up to about 25\%; for consistency, we use henceforth the Debye model values.}

The Eliashberg function can be constructed as 
$\alpha^{2}(\Omega) F(\Omega) {\simeq} \overline{\alpha}^{2} F(\Omega) \simeq \overline{\alpha}^{2} \beta\, \Omega^2$, which identifies
\begin{equation}
b = \overline{\alpha}^{2}\,\beta.
\label{eq:b_from_alpha2_beta}
\end{equation}
In Ref.~\cite{kaplan} the parameter $b$ is estimated from tunneling measurements, 
but their decay rates contain a renormalization factor $Z_1(0)$;
thus our results for $b$ should be compared to their $b/Z_1(0)$ ratio. We find reasonable agreement, with the largest deviation being $\sim {40}\%$ for Al.  Using our values for $b$ and those for the Debye energy in the fourth row of Table~\ref{tab:eph_prefactors}, we finally obtain $\tau_\mathrm{el}$ in the fifth row.

We briefly note that a more direct estimate for $b$ can in principle be obtained by measuring the power $P$ dissipated at low temperature by heated electrons, which is expected to be given by $P=\Sigma V(T_\mathrm{e}^5-T_{\mathrm{ph}}^5)$, with $V$ the sample volume, $T_\mathrm{e}$ the electronic temperature, and $T_\mathrm{ph}$ the phonon bath one. Then $b={\hbar}\Sigma/48\zeta(5)\pi k_B^5 \nu_F$, with $\zeta$ the Riemann zeta function and $\nu_F$ the electron density of states at the Fermi energy~\cite{Giazotto}. The values for $b$ estimated using experimental values for $\Sigma$~\cite{Giazotto} deviate from ours by less than one order of magnitude. We stress that there is significant uncertainty in these results, since the quadratic dependence on $\Omega$ in Eq.~\eqref{eq:A2F_quadratic} and correspondingly the fifth power in temperature in the formula for $P$ are affected by disorder~\cite{Grimvall,Giazotto,sergeev2000breakdown} {(see also Appendix~\ref{app:dirty_eph})}.

The time scale characterizing the phonon collision integral is

\begin{equation}
\tau_{\rm ph}
= \frac{\hbar N}{2\pi\,\nu_F\,\overline{\alpha}^{2}\,\Omega_D},
\label{eq:tau_ph_def}
\end{equation}
with $N$ the atomic number density and $\nu_F$ the electronic density of states at the Fermi level, see Table~\ref{tab:eph_prefactors}. Note that the ratio between the two collision times is independent of the electron-phonon coupling strength, since it can be written in the form
\begin{equation}
\frac{\tau_\mathrm{el}}{\tau_\mathrm{ph}}=\frac{\nu_F}{\beta N \Omega_D^2}.
\label{eq:tau_ratio}
\end{equation}
For superconducting metals, $\tau_\mathrm{ph}$ can be expressed in terms of the parameter $\tau_{0}^{\mathrm{PB}}$ of Ref.~\cite{kaplan1976}, 
$\tau_{\mathrm{ph}}= \tau_{0}^{\mathrm{PB}} \pi\Delta /\Omega_{D}$~\footnote{This relation should be taken as an approximate one, since it is exact only under the assumption that our effective coupling $\overline{\alpha}^2$ takes the same value as the spectral average $\langle\alpha^{2}\rangle$ of Ref.~\cite{kaplan1976}.}.

\begin{table}[!tb]
\centering
\resizebox{\columnwidth}{!}{%
  \begin{tabular}{@{}l|ccccccc@{}}
    \toprule
        & Al & Nb & Ti & Cu & Au & Ag & Ta \\
    \midrule
    $\lambda_{\rm m}$
        & 0.43 & 0.80 & 0.40 & 0.14 & 0.14 & 0.10 & 0.70 \\[2pt]
    $\overline{\alpha}^{2}$ [meV]
        & 1.75 & 2.26 & 1.37 & 0.43 & 0.28 & 0.21 & 1.53 \\[2pt]
    $10^{4}\beta$ [meV$^{-3}$]
        & {1.81} & {6.50} & {1.82} & {3.29} & {32.1} & {11.7} & {7.46} \\[2pt]    
    $10^{4}\beta_{{\mathrm{fit}}}$ [meV$^{-3}$]
        & 1.85 & 8.42 & 1.46 & 3.34 & 30.9 & 9.69 & 9.21 \\[2pt]
    $10^{4}b$ [meV$^{-2}$]
        & {3.17} & {14.7} & {2.49} & {1.41} & {8.99} & {2.46} & {11.4} \\[2pt]
    $\Omega_D$ [meV]
        & 36.45 & 23.87 & 36.71 & 29.47 & 14.22 & 19.65 & 22.23 \\[2pt]
    $\tau_{\mathrm{el}}$ [fs]
        & {6.83} & {5.24} & {8.49} & {28.9} & {40.5} & {56.2} & {8.35} \\[2pt]
    $\nu_F$ [eV$^{-1}$\,nm$^{-3}$]
        & {23.25} & {26.85} & {25.11} & {18.08} & {16.03} & {15.98} & {26.79} \\[2pt]
    $N$ [nm$^{-3}$]
        & {60.22} & {55.64} & {56.85} & {84.91} & {59.20} & {58.66} & {55.22} \\[2pt]
    $\tau_{\mathrm{ph}}$ [ps]
        & {4.25} & {4.02} & {4.72} & 38.8 & {97.2} & 93.2 & {6.35} \\
    $\Delta$ [meV]
        & 0.18 & 1.54 & 0.06 & -- & -- & -- & 0.71 \\[2pt]
    \bottomrule
  \end{tabular}%
}
\caption{Electron--phonon interaction parameters.
Mass enhancement $\lambda_{\rm m}$ from Ref.~\cite{GGrimvall_1976}. 
Effective coupling $\overline{\alpha}^2$ from Eq.~\eqref{eq:a2bar_def} using phonon density of states from Ref.~\cite{LandoltBoernstein}. 
Phonon DoS coefficient $\beta$ {($\beta_\mathrm{fit}$)} from {the Debye model, Eq.~\eqref{eq:betaD}} (a quadratic fit, see text).
Eliashberg coefficient $b = \overline{\alpha}^2 \beta$ from Eq.~\eqref{eq:b_from_alpha2_beta}. 
Debye energy $\Omega_D$ from Ref.~\cite{Gladstone}. 
Electron collision time $\tau_\mathrm{el}$ from Eq.~\eqref{eq:tau_el_def}.
Density of states at the Fermi energy $\nu_F$ {and atomic number density $N$} from {the second and third columns of Table~VI in} Ref.~\cite{Gladstone} (using the free-electron model {for $\nu_F$}).
Phonon collision time $\tau_\mathrm{ph}$ from Eq.~\eqref{eq:tau_ph_def}.
Superconducting gaps $\Delta$ are taken from Ref.~\cite{Carbotte} (Al, Nb, Ta); for Ti we use the BCS relation $\Delta = 1.764\,k_B T_c$ with $T_c$ from Ref.~\cite{Gladstone}.
}
\label{tab:eph_prefactors}
\end{table}

\subsection{Phonon injection}\label{sec:ph_inj}

In closing this section, we discuss briefly the phonon generation term $g(\Omega,t)$ in Eq.~\eqref{eq:sist_eq}. Its energy and time dependencies are determined by the particular generation mechanism under consideration, 
such as energy deposits by particles like cosmic muons or by gamma rays;
simulations of these processes are possible~\cite{wilen_etal,Modeling}. Here we consider a constant-in-time injection in a narrow band,
\begin{equation}\label{eq:g_delta}
    g(\Omega) = \tilde{g} \,\delta(\Omega - \Omega_{\text{inj}}),
\end{equation}
with $\Omega_{\text{inj}}$ the injection energy and $\tilde{g}$ the injection rate times the bandwidth. On one hand, this assumption simplifies the treatment that we present in the next sections, encoding the complex generation process into just these two free parameters. On the other hand, such generation can be approximately realized experimentally by biasing a Josephson junction at a voltage just above the pair-breaking threshold, $V_b = 2\Delta/e$~\cite{iaia_etal,Modeling,Eisenmerger_1981}; phonon injection is also possible using NIS junctions~\cite{patel.2017}. 

While as mentioned above we will treat $\tilde{g}$ and $\Omega_\mathrm{inj}$ as free parameters, as a reference point let us estimate $\tilde{g}$ for SIS injection.
At the threshold bias, the current through the junction is the critical current $I_c= \pi\Delta/ 2e R_n$, with $R_n$ the (normal-state) junction resistance. After passing through the junction, two electrons can recombine by emitting a phonon of energy $2\Delta$ at rate $I_c/(2e)$. For such a biased junction, we can estimate the injection rate $\tilde{g}$ by noting that the rate of phonon generation $\mathrm{I}_\mathrm{ph}$ is

\begin{equation}
    \mathrm{I}_\mathrm{ph} =  V N \int_{0}^{\Omega_{D}} d\Omega \, F(\Omega)\, g(\Omega) = V N F(\Omega_\mathrm{inj}) \tilde{g} \, ,
\end{equation}
with $V$ the sample volume (we assume that after recombination, all the generated phonons are transmitted to the substrate). Therefore we find
\begin{equation}
    \tilde{g} = \frac{\pi \Delta}{(2e)^2 R_n V N F(2\Delta)}.
\end{equation}
For an order-of-magnitude estimate, we consider (cf. Ref.~\cite{Modeling}) an Al junction [$\Delta= 0.18\,$meV, $F(2\Delta) = 2.4\times10^{-5}\,$meV$^{-1}$] with $R_n\sim h/(2e)^2$ of order of the (superconducting) resistance quantum on a chip of volume $V = 8\times8\times 0.5\,$mm$^3$ with Si substrate ($N= 50\,$nm$^{-3}$), finding $\tilde{g} \sim 10^{-5}\,$meV/s.
Having obtained estimates for the parameters in the model, we next focus on finding steady-state solutions for the distribution functions.

\section{Steady-state distributions}\label{sec:distributions}

Solving the kinetic equations is in general a non-trivial task: even in the steady state, Eqs.~\eqref{eq:f_st} and \eqref{eq:k_eq_eff} form a nonlinear system of coupled integral equations. While, to our knowledge, no closed form solution is available, we analytically derive certain relations and approximate results that can guide us in the numerical solution of the system.

\subsection{Analytical considerations}\label{sec:analytic}

Equation~\eqref{eq:k_eq_eff} being linear in $n(\Omega)$, we can express the latter in terms of $n_\mathrm{inj}$ and $f$ as
\begin{align}
    n(\Omega) & \simeq \frac{1}{1+\Omega/\tilde\Omega} n_\mathrm{inj}(\Omega) + n_1(\Omega) ,\label{eq:nlO} \\
    n_1(\Omega) & = \frac{1}{\Omega + \tilde\Omega} \bigg\{ 2\int_0^{\Omega_D}\!dE\, f(E+\Omega)\left[1-f(E)\right] \nonumber \\ &+ \int_0^{\Omega}\!dE\, f(E) f(\Omega-E)\bigg\}, \label{eq:n1}
\end{align}
where
\begin{equation}\label{eq:tildeO}
 {\tilde{\Omega} = \tau_\mathrm{ph} \Omega_D/\bar{\tau}_{ms}}
\end{equation}
and as only approximation we assumed that $f$ decays quickly above an energy small compared to $\Omega_D$ (or equivalently, that we can set $f(E)=0$ for $E>\Omega_D$).

Specializing now to the case of monocromatic injection, we consider two limiting cases.
Although as we discuss below not relevant in practice, we consider first the limit of fast phonon escape from the metal $\tau_{ms} \to 0$, which implies $\bar\tau_{ms} \to 0$ and $\tilde\Omega \to \infty$. In this regime, the phonon distribution reduces to
\begin{equation}\label{eq:n0inj}
n_0(\Omega) = \tilde{n}_{\rm inj}\,\delta(\Omega-\Omega_{\rm inj}),
\end{equation}
where $\tilde{n}_\mathrm{inj}$ is related to $\tilde{g}$ in the same way as $n_\mathrm{inj}$ is related to $g$ by Eq.~\eqref{eq:ninj}. Then
the electron kinetic equation [see Eqs.~\eqref{eq:Iel} and \eqref{eq:f_st}]
can be written as
\begin{align}
0  = &
\int_0^{\Omega_D}d\Omega\,\Omega^2 f_{E+\Omega}(1-f_E)
-\int_E^{\Omega_D} d\Omega \, \Omega^2f_{E}f_{\Omega-E}
\nonumber \\
-& \int_0^{E}d\Omega\,\Omega^2 f_{E}(1-f_{E-\Omega})  
\nonumber \\
+&\tilde{n}_{\rm inj}\Omega_{\rm inj}^2 \theta(\Omega_\mathrm{inj}-E) \left[1 - 2f_{E} - f_{\Omega_\mathrm{inj}-E} + f_{E+\Omega_\mathrm{inj}} \right]
\nonumber \\
+&\tilde{n}_{\rm inj}\Omega_{\rm inj}^2 \theta(E-\Omega_\mathrm{inj}) \left[f_{E+\Omega_\mathrm{inj}} + f_{E-\Omega_\mathrm{inj}} -2f_{E}\right], 
\label{eq:schematic_electron}
\end{align}
where for notational compactness the argument of the distribution functions is written as a subscript. This equation can be rendered dimensionless by rescaling the energies by
\begin{equation}
\bar E = \left(\tilde{n}_\mathrm{inj}\,\Omega_\mathrm{inj}^2\right)^{1/3}.
\label{eq:Ebar_def}
\end{equation}
For $\bar{E},\, \Omega_\mathrm{inj} \ll \Omega_D$, we expect a one-parameter family of solutions
\begin{equation}\label{eq:fastescf}
f(E) = \mathcal{F}\!\left(\frac{E}{\bar E}; \frac{\Omega_\mathrm{inj}}{\bar E}\right),
\end{equation}
with $\mathcal{F}(0;y) = 1/2$ and $\mathcal{F}(x;y) \sim e^{-2x^{5/2}/(5\sqrt{3}y)}$ for $x\gg 1,\, y$~\footnote{The behavior of $\mathcal{F}(x;y)$ at large $x$ is found by noticing that assuming $f$ to decay sufficiently fast for $E\gg \bar{E},\, \Omega_\mathrm{inj}$, the dominant terms in Eq.~\eqref{eq:schematic_electron} are the second line (where $f_{E-\Omega}$ can be neglected compared to unity) and the last line. In the latter, we can expand to second order in $\Omega_\mathrm{inj}$ and obtain for $\mathcal{F}$ a generalized Airy equation of the form (cf. Ref.~\cite{Paul})
\begin{equation}
    y^2 \mathcal{F}''(x;y) - \frac{x^3}{3} \mathcal{F}(x;y) = 0
\end{equation}
The solution to this equation is a modified Bessel function of the second kind, from which we obtain the asymptotic expression given in the main text.} 
In fact, for any value of $\bar\tau_{ms}$ we always have $f(0)=1/2$ and
\begin{equation}\label{eq:n0}
    n(0) = 2 \frac{\bar\tau_{ms}}{\tau_\mathrm{ph}\Omega_D} \int_0^{\Omega_D}\!dE \, f(E) \left[1-f(E)\right] \simeq 
  {\frac{\bar{E}}{2\tilde{\Omega}}},
\end{equation}
where in the last estimate we assumed $f$ to be bounded by $1/2$ and to decay sufficiently fast above an energy of order $\bar{E}$.

When $\bar\tau_{ms}$ is small but non-zero, we can proceed iteratively and write the phonon distribution function in the approximate form
\begin{equation}
    n(\Omega) \simeq n_0(\Omega) + \bar{\tau}_{ms} I_\mathrm{ph} \left\{n_0, f\right\}.
\end{equation}
Rescaling the energy as above, we see that in the last term two dimensionless quantities appear, $\bar{E}/\tilde{\Omega}$ and $\tilde{n}_\mathrm{inj}/\tilde{\Omega}$. 
The smallness of the first parameter ensures a small phonon density in the metal [cf. Eq.~\eqref{eq:n0}], as it implies that phonons escape from the metal faster than they are scattered by the electrons. More generally, the smallness of the two quantities determines the validity of the iterative approximation, so deviations of the solution from the above one-parameter form will become apparent and grow as the largest of them increases beyond being of order unity. This criterion shows that the fast phonon escape approximation is not important in practical cases: if $\bar{E}$ is much less than the gap, we expect that the metal can be efficient at down-converting the injected phonons (as we discuss in the next section); therefore, let us assume an injection rate such that $\bar{E}$ is one order of magnitude smaller than the gap in Al; then we can estimate $\bar{E}/\Omega_D\sim 10^{-3}$. Using Eq.~\eqref{eq:effective_parameters} and our parameter estimates, we have $\bar\tau_{ms}\sim 1$-$10\,\mu$s and therefore $\bar{E}/\tilde{\Omega} \sim 10$-$10^{3}\gg 1$.

In the opposite limit to that considered above (that is, for $\bar\tau_{ms} \to \infty$ and $\tilde\Omega\to 0$), the solution takes the equilibrium-like form
\begin{equation}\label{eq:approx_distributions}
    f(E) = \frac{1}{e^{E/\tilde{E}}+1}\,,\quad n(\Omega) = \frac{1}{e^{\Omega/\tilde{E}}-1}\, ,
\end{equation}
where the energy scale $\tilde{E}$ plays the role of an effective temperature and must be determined by some additional consideration. If we treat the left-hand side of Eq.~\eqref{eq:k_eq_eff} as the actual collision integral for $n$, requiring the balance between the power injected in the phonon system and the one lost by escape of phonon from the metal results in the condition
\begin{equation}\label{eq:Etilde_cond}
    \int_0 \Omega^3 n(\Omega) d\Omega = \int_0 \Omega^3 n_\mathrm{inj}(\Omega) d\Omega \, ,
\end{equation}
which gives $\pi^4 \tilde{E}^4/15 = \tilde{n}_\mathrm{inj}\Omega_\mathrm{inj}^3$ [by dropping the upper integration limit, we are ignoring a small correction to the numerical coefficient of order $\sim (\Omega_D/\tilde{E})^3 e^{-\Omega_D/\tilde{E}}$]; this suggests defining
\begin{equation}\label{eq:tilde_E}
 \tilde{E} = \frac{15^{1/4}}{\pi}\left(\tilde{n}_\mathrm{inj}\Omega_\mathrm{inj}^3\right)^{1/4} 
\end{equation}
in terms of which we obtain [cf. Eq.~\eqref{eq:n0}] $n(0) = \tilde{E}/\tilde{\Omega}$. %, 
In fact, 
substituting Eq.~\eqref{eq:approx_distributions} for $f$ into Eq.~\eqref{eq:n1}, at finite $\bar\tau_{ms}$ and for $\Omega < \Omega_\mathrm{inj}$ we expect a better approximation for $n_1$ to be
\begin{equation}\label{eq:n_approx}
    n_1(\Omega) = \frac{1}{1+{\tilde{\Omega}/\Omega}}
    \, \frac{1}{e^{\Omega/\tilde{E}}-1} \, .
\end{equation}
This expression, {whose value as $\Omega\to 0$ is the one estimated above}, shows that deviation from the equi\-librium-like form should be negligible when $\tilde\Omega 
\ll \tilde{E}$. 

Note that when $\tilde{\Omega}/\Omega_\mathrm{inj}\ll 1$, the peak in the phonon occupation at $\Omega_\mathrm{inj}$ is strongly suppressed by this small factor compared to the injected value [cf. Eq.~\eqref{eq:nlO}]. 
We next compare these expectations to the results of numerical calculations.

\subsection{Numerical solution}
\label{sec:numerical_results}

We solve the coupled stationary kinetic equations for electrons and phonons in the metal,
Eqs.~\eqref{eq:f_st} and~\eqref{eq:k_eq_eff}, using the damped Newton--Raphson scheme described in Appendix~\ref{app:numerical}. We consider monochromatic injection in the substrate,
Eq.~\eqref{eq:g_delta}, and we take $\tau_\mathrm{es}=50~\mu{\rm s}$ as a reference value. Since we target Al as the superconductor, we consider $\Omega_\mathrm{inj} > 2\Delta_\mathrm{Al} = 0.36\,$meV, and as the energies involved should be small compared to $\Omega_D$ in Table~\ref{tab:eph_prefactors}, we restrict $\Omega_\mathrm{inj} \leq 12 \Delta_\mathrm{Al}=2.16\,$meV. Moreover, as discussed in Appendix~\ref{app:eeint}, we need $\tilde{g} \gtrsim {10^{-2}}\,$meV/s to be able to neglect electron-electron interaction and thermal phonon effects, while restricting $\tilde{E}$ of Eq.~\eqref{eq:tilde_E} to a similar upper bound as for $\Omega_\mathrm{inj}$ gives $\tilde{g} \lesssim 10^7\,$meV/s. The corresponding range for $\tilde{E}$ goes from $\sim10^{-2}\,$meV to a few meV.

Numerical solutions are computed on a {$5\times40$} grid in
{injection energy times injection rate}.
Specifically, the injection rate $\tilde{g}$ is sampled at
{40} equally spaced values (on a logarithmic scale) between
$\tilde{g}_{\min}={10^{-2}}~\mathrm{meV/s}$ and
$\tilde{g}_{\max}=10^{7}~\mathrm{meV/s}$.
The injection energy $\Omega_{\rm inj}$ is sampled at
{5} uniformly spaced values in the interval
$[\,2\Delta_{\rm Al}+{0.01}\,{\mathrm{meV}},
\,12\Delta_{\rm Al}\,]$.
For each metal-substrate pair, we use as transmission lifetimes $\tau_{ms}$ and $ \tau_{sm}$ the average of those computed from the AMM and DMM models, see Table~\ref{tab:tau_ell}.
These are then mapped to the effective escape time $\bar\tau_{ms}$ and to the effective injected phonon distribution $n_{\rm inj}(\Omega)$ through Eqs.~\eqref{eq:effective_parameters} and \eqref{eq:ninj}. Note that from the values in Tables~\ref{tab:tau_ell} and \ref{tab:eph_prefactors} we estimate $\tilde\Omega \sim 10^{-4}\,$meV$\,\ll 2\Delta_\mathrm{Al}$, ensuring strong suppression of the injected population at $\Omega_\mathrm{inj}$ as well as small deviations from the equilibrium-like distribution for phonons, since $\tilde\Omega \ll \tilde{E}$.

\begin{figure}[tb]
    \centering
    \includegraphics[width=0.99\columnwidth]{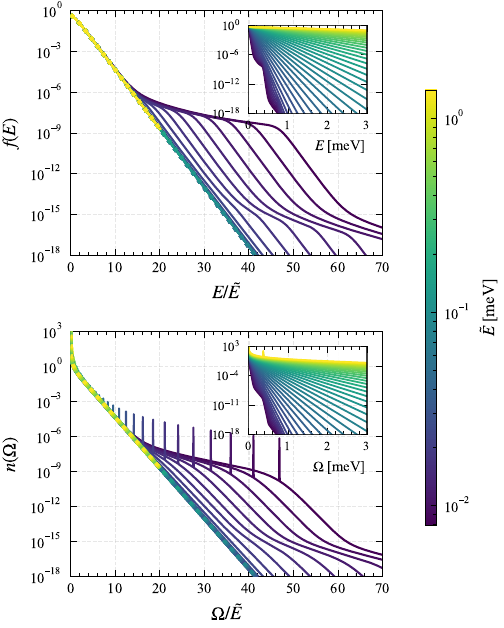}
    \caption{
    Electron distribution $f(E)$ {(top panel)} and phonon distribution $n(\Omega)$ {(bottom)}
    in a Cu film under steady, monochromatic phonon injection at $\Omega_{\rm inj} = 0.37 \, \mathrm{meV} \gtrsim 2\Delta_{\rm Al}$. 
    The dashed lines correspond to the electron distribution in Eq.~\eqref{eq:approx_distributions} in {the top
    panel} and to the phonon distribution in Eq.~\eqref{eq:n_approx} in {the bottom one}.
    }
    \label{fig:Cu_distributions}
\end{figure}

To illustrate the structure of the stationary solutions, in Fig.~\ref{fig:Cu_distributions} we plot the electron and phonon distributions in a Cu film for a fixed injection energy $\Omega_{\rm inj} =0.37\,\mathrm{meV} \gtrsim 2\Delta_{\rm Al}$ and different injection rates; similar plots are obtained for higher injection energies and, at energies below $\Omega_\mathrm{inj}$, for injection with a Gaussian spectrum, see Appendix~\ref{app:more_sol}.
In the main panels the energy axes are normalized by $\tilde E$, and at low energies, $E,\,\Omega \lesssim 12\tilde{E}$, for all the considered values of $\tilde{E}$ the numerical solutions are well approximated by Eqs.~\eqref{eq:approx_distributions} and \eqref{eq:n_approx} for $f$ and $n$, respectively. In fact, when $\Omega_\mathrm{inj}/\tilde{E} \lesssim 16$, the approximate formulas remain applicable to higher energies and all the curves approximately collapse into a single one. For $\Omega_\mathrm{inj}/\tilde{E} \gtrsim 16$ the curves display, on logarithmic scale, an exponential fall followed by a slower decrease that {extends} up to $\Omega\sim \Omega_\mathrm{inj}$, this structure qualitatively repeating at integer multiples of $\Omega_\mathrm{inj}$. The {`step'} structure originates from the absorption of injected phonons by low-energy electrons and their subsequent relaxation by phonon emission, similar to the regime of `cold' quasiparticles absorbing photons studied in Ref.~\cite{basko.2019}.
{While we discuss analytical approximations for the first steps in Appendix~\ref{app:steps}, we do not explore this structure in more detail here,} as the knowledge of the distributions at low energies is sufficient for estimating the down-conversion efficiency, which we consider next.

\section{Down-conversion efficiency}
\label{sect:DW_eff}

To quantify how effectively the normal metal suppresses quasiparticle generation in a superconducting device, we need to consider the generation rate~\footnote{{We consider here the generation rate for the quasiparticle density normalized by the Cooper pair density \(x_\mathrm{qp}={\mathcal{N}}_\mathrm{qp}/\nu_F \Delta\) with \({\mathcal{N}}_\mathrm{qp}\) of Eq.~\eqref{eq:Nqp_def}.}} due to pair-breaking phonons~\cite{Paul}
\begin{align}
g_{qp}
&=
\frac{{2}}{\tau_0 T_c^3{\Delta}}
\int_{\Delta}\! dE
\int_{E+\Delta}^{\Omega_D}\! d\Omega \;\Omega^2
\nonumber \\
&\quad\times
\frac{E(\Omega-E)+\Delta^2}
{\sqrt{E^2-\Delta^2}\,\sqrt{(\Omega-E)^2-\Delta^2}}
\,n_S(\Omega)
\nonumber \\
&=
\frac{{2}}{\tau_0 T_c^3}
\int_{2\Delta}^{\Omega_D}\! d\Omega \;\Omega^2
S_+\!\left(\frac{\Omega}{\Delta}\right)n_S(\Omega),
\label{eq:gqp_full}
\end{align}
where \(\tau_0\) is proportional to \(\tau_{\mathrm{el}}\) (see Sec.~\ref{sec:E-PH}), \(T_c\) is the critical temperature, {\(n_S\) the phonon distribution in the superconductor}, and \(S_+\) is the spectral density {for pair-breaking processes}~\cite{LectNotes} in the regime of low quasiparticle density (see Appendix~\ref{app:ns_nS_relation} {for the concrete expression}). We define the down-conversion efficiency as
\begin{equation}
R_{qp}\equiv 1-\frac{g_{qp}}{g_{qp}^{\mathrm{nm}}},
\label{eq:Rqp_definition}
\end{equation}
where \(g_{qp}^{\mathrm{nm}}\) is the `no-metal' quasiparticle generation rate in the absence of the metallic film. By construction, \(R_{qp}=1\) means complete suppression of quasiparticle generation, \(R_{qp}=0\) implies that the metal has no effect, and \(R_{qp}<0\) indicates that the metal is actually detrimental.

The phonon distribution \(n_S\) is {approximately} related to that in the substrate by (Appendix~\ref{app:ns_nS_relation})
\begin{equation}
n_S(\Omega)
=
n_s(\Omega)\,
\frac{1}{1+\tau_{Ss}/\tau_{\mathrm{pb}}(\Omega)},
\label{eq:nS_of_ns_general_main}
\end{equation}
with
\begin{equation}
\frac{1}{\tau_{\mathrm{pb}}(\Omega)}
=
\frac{1}{\pi\tau_{0}^{\mathrm{PB}}}
S_+\!\left(\frac{\Omega}{\Delta}\right)
\label{eq:pb_rate_def_main}
\end{equation}
being the inverse lifetime of phonons with energy $\Omega>2\Delta$ due to pair breaking, and $\tau_{Ss}$
the phonon transmission times between superconductor ($S$) and substrate ($s$), analogous to $\tau_{ms}$. 
The last factor in the denominator of Eq.~\eqref{eq:nS_of_ns_general_main} accounts for the so-called ``phonon trapping'' effect~\cite{Chang.1978,Paul}. In deriving Eq.~\eqref{eq:nS_of_ns_general_main} we assumed that quasiparticles in the superconductor remain cold, $\tilde{E}\lesssim \Delta$.

For the `no-metal' case, using {Eqs.~\eqref{eq:phonon_substrate_no_metal} and \eqref{eq:nS_of_ns_general_main} in Eq.~\eqref{eq:gqp_full}}, the generation rate is
\begin{equation}
    g_{qp}^{\mathrm{nm}} = \frac{2\Omega_\mathrm{inj}^2}{\tau_0 T_c^3}  
    \frac{S_+(\Omega_\mathrm{inj}/\Delta)}{1+\lambda S_+(\Omega_\mathrm{inj}/\Delta)}\tau_\mathrm{es}\tilde{g} \, ,
\end{equation}
where 
\begin{equation}
\lambda = \frac{\tau_{Ss}}{\pi\tau_0^\mathrm{PB}}.     
\end{equation}
For a 100~nm thick Al film on Si, we estimate \(\lambda \simeq {0.42}\) [cf. Tables~\ref{tab:tau_ell} and \ref{tab:eph_prefactors} {and Appendix~\ref{app:escape}}]. Not surprisingly, faster phonon escape from the substrate (shorter $\tau_\mathrm{es}$) 
{suppresses} generation.
In the ideal case, the metal would act as a perfect absorber, in which case [cf. Eq.~\eqref{eq:perf_absorp}] we find \(g_{qp}^\mathrm{pa} = g_{qp}^\mathrm{nm}\tau_{sm}/(\tau_{sm}+\tau_\mathrm{es})\) and the down-conversion efficiency is
\begin{equation}
    R_{qp}^\mathrm{pa} = \frac{1}{1+\tau_{sm}/\tau_\mathrm{es}} .
\end{equation}
This expression shows that, for a given escape time a necessary condition to achieve high efficiency is \(\tau_{sm} \ll \tau_\mathrm{es}\), since otherwise escape from the substrate to the sample holder dominates over down-conversion in the normal metal. This condition is in fact satisfied for typical parameters, see Table~\ref{tab:tau_ell}, from which we estimate \(R_{qp}^\mathrm{pa} \simeq 98.5\)-\(99\%\).

Using Eq.~\eqref{eq:phonon_substrate} for \(n_s(\Omega)\) in Eqs.~\eqref{eq:gqp_full}-\eqref{eq:nS_of_ns_general_main} we arrive at
\begin{align}\label{eq:Rqp_n}
R_{qp}
= R_{qp}^{\mathrm{pa}} & \left[1
-  \frac{1}
{\tilde g{\tau_{\mathrm{es}}}\Omega_{\mathrm{inj}}^2} {\frac{1+\lambda S_+(\Omega_\mathrm{inj}/\Delta)}{\lambda S_+(\Omega_\mathrm{inj}/\Delta)}} \right.
\\ & \left. \times\int_{2\Delta}^{\Omega_D}\! d\Omega\;\Omega^2 {\frac{\lambda S_+(\Omega/\Delta)}{1+\lambda S_+(\Omega/\Delta)}} n(\Omega)\right],
\nonumber
\end{align}
{proving that in fact the perfect absorption limit gives an upper bound for the down-conversion efficiency. Using the thermal-like form for $n$, Eq.~\eqref{eq:approx_distributions}, applicable for \(\tilde\Omega\ll\tilde{E}\), a good approximation for the efficiency when \(\lambda \gtrsim 0.5\) is (see Appendix~\ref{app:Rqp_minimum})}
\begin{equation}\label{eq:RqpRlambda}
 R_{qp} \simeq R_{qp}^{\mathrm{pa}}\left[1-\frac{15}{\pi^4}\frac{\Omega_\mathrm{inj}}{\Delta}\mathcal{R}_\lambda\left(\frac{\tilde{E}}{\Delta}\right)\right] ,
\end{equation}
where
\begin{equation}\label{eq:Rlambda}
   \mathcal{R}_\lambda\left(x\right) \simeq A_\lambda e^{-2/x}\left(\frac{4}{x^3}+\frac{4}{x^2} +\frac{1+\pi\lambda}{\pi\lambda}\frac{2\zeta(3)}{x}\right)
\end{equation}
with
\begin{equation}
    A_\lambda = \frac{1+\lambda S_+(\Omega_\mathrm{inj}/\Delta)}{\lambda S_+(\Omega_\mathrm{inj}/\Delta)} \frac{\pi\lambda}{1+\pi\lambda} \, .
\end{equation}
The above formulas show that for $\tilde{E}\ll \Delta$ the efficiency is given by the perfect absorption result up to exponentially small corrections. As $\tilde{E}$ approaches $\Delta$, the efficiency decreases, and the decrease is stronger the higher the injection frequency. For $\tilde{E} > \Delta$ the efficiency seems to recover, but this is likely an artifact due to the approximations used to derive Eq.~\eqref{eq:nS_of_ns_general_main} not being valid anymore; we therefore exclude this region from further consideration. 

\begin{figure}[tb]
\centering
\includegraphics[width=\columnwidth]{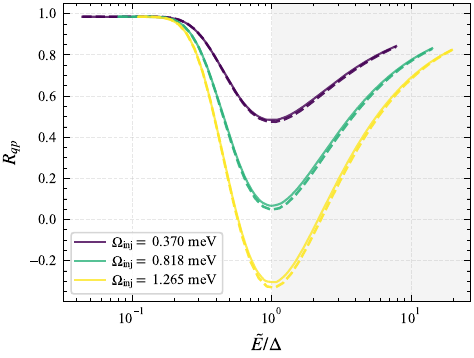}
\caption{
Down-conversion efficiency $R_{qp}$, Eq.~\eqref{eq:Rqp_definition}, vs. normalized `effective temperature' $\tilde{E}/\Delta$ for a 100~nm Al film ($\Delta=0.18\,$meV, $\lambda\simeq {0.42}$) and three values of injection energy, see legend. The solid lines are obtained from numerics, the dashed lines using Eq.~\eqref{eq:RqpRlambda} (the discrepancy is due to $\lambda < 0.5$ being outside the validity regime of analytical approximations). In the shaded region, $\tilde{E}/\Delta > 1$, energetic quasiparticles in the superconductor can emit phonons and limit the efficiency; we exclude this regime from our analysis. 
}
\label{fig:Rqp_vs_delta}
\end{figure}

To illustrate the {above} findings, we plot in Fig.~\ref{fig:Rqp_vs_delta} the down-conversion efficiency $R_{qp}$ as a function of $\tilde E/\Delta$ for {a 100-nm thick Al film with $5\,\mu$m-thick Cu as the normal metal, calculated for} three %fixed
values of $\Omega_{\mathrm{inj}}$. The analytical approximation of Eq.~\eqref{eq:RqpRlambda} (dashed curves) is in fair agreement with the numerics (solid), even though for this Al thickness we have $\lambda \simeq {0.42}<0.5$. {For a different normal metal, the transmission coefficients and hence the transmission times differs from those of Cu, see Tables~\ref{tab:avg_coefficients} and \ref{tab:tau_ell}; however, these differences only affect the prefactor $R_{qp}^{\mathrm{pa}}$ and, as mentioned above, only very slightly. As a consequence, the curves obtained for other metals are indistinguishable from those for Cu in the used scale.}

So far we have considered a relatively thick normal metal film, $d_m=5\,\mu$m, for which
high efficiency is maintained for $\tilde{E} \lesssim 0.2 \Delta$, see Fig.~\ref{fig:Rqp_vs_delta}. {This finding is valid} under the assumption $\tilde{\Omega} \ll \tilde{E}$, a condition that can be violated for thin metallic films since $\tilde{\Omega}/\tilde{E} \propto {d_s/d_m}$ {so long as $\tau_{sm} \ll \tau_\mathrm{es}$}. Requiring $\tilde{\Omega} \lesssim 0.1 \tilde{E}$, using the definitions in Eqs.~\eqref{eq:effective_parameters} and \eqref{eq:tildeO}, {and assuming $\tau_\mathrm{es} \gg \tau_{sm}$,} we arrive at $\tau_{ms} \gtrsim 50 (\tau_{sm}/\tau_\mathrm{es}) (\tau_\mathrm{ph} \Omega_D/\Delta)$, or equivalently [cf. Eq.~\eqref{eq:transmission_times}]
\begin{equation}\label{eq:d_m_min}
    d_m \gtrsim 50 d_s \frac{c_m \eta_{ms}}{{p} c_s \eta_{sm}} \frac{\tau_\mathrm{ph} \Omega_D}{\tau_\mathrm{es}\Delta} \, ,
\end{equation}
{where we include explicitly the normal-metal coverage fraction $p$. This inequality ensures} that at low injection rate the efficiency is suppressed at most by a few percent, the suppression being of order $\tilde\Omega/\Delta$. For Cu as the normal metal {with $p=0.64$}, Al as the superconductor, and 0.5~mm thick Si as the substrate, with $\tau_\mathrm{es} = 50\,\mu$s this results in $d_m \gtrsim {450}\,$nm as the minimum thickness for efficient down-conversion. {According to Eqs.~\eqref{eq:d_m_min} and \eqref{eq:etaDMM}, the metal thickness can be reduced using a material with slower speed of sound; in fact,} {considering Au and Ag, the minimum thicknesses are 80 and 290~nm, respectively; the values for Al$_2$O$_3$ as substrate are about 57\% of those for Si. We caution that these estimates are dependent on the assumed value for the substrate phonon escape time $\tau_\mathrm{es}$, which as discussed in Sec.~\ref{sec:Phonon escape time} is not well determined; nonetheless, they indicate that Au might be preferable to Cu for phonon down-conversion.} 

\begin{figure}[tb]
\centering
\includegraphics[width=\columnwidth]{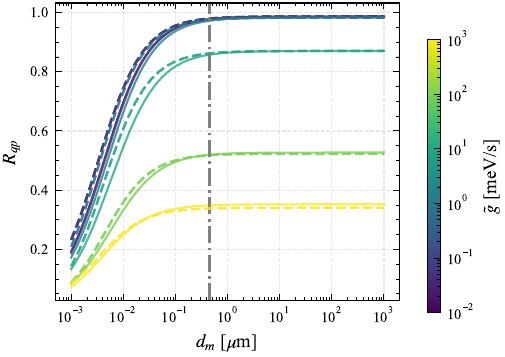}
\caption{
Down-conversion efficiency $R_{qp}$ vs. normal-metal thickness for $\Omega_\mathrm{inj}=0.5\,$meV; the metal is Cu, other parameters as in Fig.~\ref{fig:Rqp_vs_delta}. Solid lines: full numerical evaluation. Dashed lines: analytical approximation of Eq.~\eqref{eq:Rqp_final}. 
{The generation rates $\tilde{g}$ used correspond to $\tilde{E}/\Delta \simeq 0.03\mathrm{-}0.98$.}
The vertical dot-dashed line demarcates the minimum thickness for efficient down-conversion according to Eq.~\eqref{eq:d_m_min}.
}
\label{fig:Rqp_vs_dm}
\end{figure}

So far we have assumed that the normal metal is thick enough so that $\tilde\Omega \ll \Omega_\mathrm{inj}$; then the peak at the injection frequency is strongly suppressed [see Eq.~\eqref{eq:nlO}] and can be ignored when estimating the down-conversion efficiency. For films thinner than the bound in Eq.~\eqref{eq:d_m_min}, since not many down-converted phonons will be present in the metal [$n(0) \simeq \tilde{E}/\tilde{\Omega} \propto {d_m/d_s}$], the efficiency is {dominated} 
by the injection peak, now only weakly suppressed, leading to
\begin{equation}\label{eq:Rqp_thin}
    R_{qp} \simeq R_{qp}^\mathrm{pa} \frac{1}{1+\tilde{\Omega}/\Omega_\mathrm{inj}}.
\end{equation}
an expression valid also in the regime $\bar{E}/\tilde{\Omega} < 1$ in which the iterative approach is applicable. As the metal thickness decreases, $\tilde{\Omega}$ increases, lowering the efficiency. 
{An approximate} expression for $R_{qp}$ {encompassing both thin and thick normal films} is obtained {by reducing the right-hand side of Eq.~\eqref{eq:RqpRlambda} by the same factor as in Eq.~\eqref{eq:Rqp_thin} to get}
\begin{equation}\label{eq:Rqp_final}
 {R_{qp} \simeq \frac{R_{qp}^{\mathrm{pa}}}{1+\tilde{\Omega}/\Omega_\mathrm{inj}}\left[1-\frac{15}{\pi^4}\frac{\Omega_\mathrm{inj}}{\Delta}\mathcal{R}_\lambda\left(\frac{\tilde{E}}{\Delta}\right)\right],}
\end{equation}
{where $\tilde{E}$ is now implicitly defined as the solution to}~\footnote{{This equation for $\tilde{E}$ interpolates between those obtained by substituting $n_1$ of Eq.~\eqref{eq:n_approx} for $n$ in the left-hand side of Eq.~\eqref{eq:Etilde_cond} and taking the limits $\tilde{\Omega}=0$ and $\tilde{\Omega} \to \infty$.} {Interestingly, in the latter limit we find $\tilde{E}^5 \propto \tilde{n}_\mathrm{inj} \Omega_\mathrm{inj}^4$, which is the energy scale determining the asymptotic behavior in the fast phonon escape regime, cf. Eq.~\eqref{eq:fastescf} and the text following it.}}
\begin{equation}
    {\tilde{E}^4 = \left[\frac{15}{\pi^4}+
    \frac{\tilde{\Omega}}{24\zeta(5)\tilde{E}}\right]\frac{\tilde{n}_\mathrm{inj}\Omega_\mathrm{inj}^3}{1+\tilde{\Omega}/\Omega_{\rm inj}.}}
\end{equation}
We compare this full expression to numerical results in Fig.~\ref{fig:Rqp_vs_dm}, where we plot efficiency vs thickness for a few injection rates. The full expression reproduces the suppression of the efficiency with decreasing thickness found numerically; moreover,
the efficiency is {approximately constant}
for thickness above the bound given in Eq.~\eqref{eq:d_m_min}, confirming its usefulness.

\section{Conclusions}\label{sec:conclusions}

In this work we have studied phonon down-conversion by a normal-metal film covering one side of an insulating substrate using a kinetic equation approach. We have revisited the determination of the parameters entering the kinetic equations, such as phonon transmission coefficients at the metal-substrate interface and the electron-phonon scattering times, see Sec.~\ref{sec:par_est}. Under steady injection of phonons, we have derived approximate formulas for the distribution functions of phonons and electrons in the metal; at sufficiently low energy they are well approximated by equilibrium-like ones, Eq.~\eqref{eq:approx_distributions}, with an effective temperature $\tilde{E}$ that depends on the rate and energy of phonon injection as in Eq.~\eqref{eq:tilde_E}. These analytical results are confirmed by numerics, see Fig.~\ref{fig:Cu_distributions}.

To quantify the impact of the down-conversion process on superconducting devices, we have introduced an efficiency parameter $R_{qp}$ [Eq.~\eqref{eq:Rqp_definition}] that compares quasiparticle generation in the superconductor in the presence of the normal metal to that in its absence. At low injection rate, such that the effective temperature is small compared to the superconducting gap, the efficiency is typically high. {Interestingly, we find that a minimum metal thickness is needed to ensure high efficiency, see Eq.~\eqref{eq:d_m_min}}.

The model considered here is much simpler than what can be simulated numerically using a Monte Carlo method~\cite{Modeling}, ignoring in particular any geometric detail beside thicknesses {and coverage fraction}. Our work is, however, complementary to such an approach in two aspects: we study the steady state, which could be challenging to reach in simulations. More importantly, our model takes into account the possibility, ignored in the simulations, that the electronic {system} heats up, leading to a decrease in down-conversion efficiency. Our estimates for phonon injection by biased junctions indicate that heating can be neglected even for continuous injection, thus validating, for such an injection, Monte Carlo simulations that treat each phonon independently and the metal as always cold.

\acknowledgments

This work is supported by the U.S. Government under ARO Grant No. W911NF-22-1-0257.

\appendix

\section{Electron-electron interaction}\label{app:eeint}

The kinetic equations in Sec.~\ref{sec:model} neglect electron–electron (e–e) interaction, assuming that electron–phonon (e–ph) processes have a dominant role in determining the electronic distribution function $f$. In this appendix we justify this approximation. 

For our purposes, we need to compare the e-ph scattering time ($\tau_\mathrm{ep}$) to the e-e one ($\tau_\mathrm{ee}$); the latter at low temperature (strictly speaking, $T=0$) for disordered metals reads~\cite{Rammer}
\begin{equation}
\tau_{\mathrm{ee}}^{-1}(E)
=\frac{E^{3/2}}{12\pi^2\sqrt{2}\,\hbar^{5/2}\,\nu_F D^{3/2}},
\label{eq:tau_ee_diffusive_recap}
\end{equation}
where $D=v_F^2\tau_\ell/3$ is the diffusion constant, with $v_F$ the Fermi velocity and $\tau_\ell$ the elastic scattering time for electrons off impurities; this expression applies for $E \tau_\ell \ll {\hbar}$ and assuming $E_F\tau_\ell \gg {\hbar}$ (with $E_F$ the Fermi energy). The time $\tau_\ell$ can be estimated from the residual resistivity $\rho_0$ using Drude's relation
\begin{equation}
\tau_{\ell}
= \frac{1}{\varepsilon_0 \omega_p^2 \rho_0}
= \frac{\mathrm{RRR}}{\varepsilon_0 \omega_p^2 \rho(300\,\mathrm{K})},
\end{equation}
with $\omega_p$ the plasma frequency and $\mathrm{RRR}=\rho(300\,\mathrm{K})/\rho_0$ the residual resistivity ratio; although the latter is typically larger than one, for our purposes it is sufficient to take it of order unity, so we henceforth set $RRR=1$. The typical room-temperature resistivity of metals is of order a few times $10^{-8}\,\Omega$m~\cite{crc} and the plasma frequency of order 10\,eV/$\hbar$~\cite{drude}; therefore we have $\tau_\ell~\sim 10^{-14}\,$s. Together with the typical Fermi velocity $v_F\sim 10^6\,$m/s~\cite{crc}, we estimate $D \sim 10^{-2}\,$m$^2/$s.

The electron-phonon scattering time is given by [cf. last term in the second line of Eq.~\eqref{eq:Iel}]
\begin{equation}
\tau_{\mathrm{ep}}^{-1}(E)
=\frac{1}{\tau_\mathrm{el}\Omega_D^3}\int_0^E\!\! d\Omega\, \Omega^2
= \frac{1}{3\tau_\mathrm{el}}\left(\frac{E}{\Omega_D}\right)^{3}.
\label{eq:tau_eph_recap}
\end{equation}
We define a cross-over energy $E_c$ by equating the two scattering times,
\begin{equation}
    E_c = \left(\frac{\tau_\mathrm{el}}{4\pi^2\sqrt{2}\hbar\nu_F}\right)^{2/3}\frac{ \Omega_D^2}{{\hbar}D} \, .
\end{equation}
Using the above estimate for $D$ and the values of $\tau_\mathrm{el}$, $\Omega_D$, and $\nu_F$ in Table~\ref{tab:eph_prefactors} {for normal metals}, we get $E_c\sim {5\text{-}16}\,\mu$eV. For energies larger than $E_c$, e-ph scattering dominates over e-e scattering, so our approximation of neglecting e-e interaction is appropriate if the energy scale {$\tilde{E}$ in Eq.~\eqref{eq:tilde_E}}
characterizing the width of the distribution function $f$ is large compared to $E_c$. Since we are interested in injection energies of order a few times {($n=2$-12)} the gap $\Delta$ (typically in Al), the requirement ${\tilde{E}} > E_c$ translates into a lower bound on the injection rate, {$\tilde{g} > {(\pi E_c)^4 /[15(n\Delta_\mathrm{Al})^3 \tau_\mathrm{es}]} \sim {10^{-5}}$-${10^{-1}}\,$meV/s.} This bound is {generally} larger than the injection rate we estimated for injection by SIS junctions in Sec.~\ref{sec:ph_inj}; therefore, that injection mechanism is unlikely to significantly drive the metal out of (quasi)equilibrium {(the possible exception being Au at high injection energy). For concreteness, we fix the lower bound $\tilde{g}> 10^{-2}\,$meV/s close to the upper end of the estimated range.}
We also note that since $E_c/k_B \sim 0.1\,$K, for $\tilde{E}> E_c$ we can ignore the thermal phonon population.

\section{Escape time estimation}\label{app:escape}

Measurements of the qubit decay rate as function of time can give information on the time evolution of excess quasiparticles~\cite{Wang2014}. Here we assume that the excess quasiparticles are due to non-equilibrium phonons in the substrate that are then transmitted into the qubit's superconductor; the dynamics of quasiparticles and phonons are governed by a set of kinetic equations similar to Eqs.~\eqref{eq:sist_eq}-\eqref{eq:keq_f}, see for instance Ref.~\cite{Paul} {(cf. Appendix~\ref{app:ns_nS_relation})}. Simpler equations of the Rothwarf-Taylor (RT) type~\cite{Rothwarf.1967} for the dynamics of excess quasiparticle density (${\mathcal{N}}_\mathrm{qp}$), pair-breaking phonon density in the superconductor (${\mathcal{N}}_S$), and in the substrate (${\mathcal{N}}_s$) can be derived from the kinetic equations by multiplying them by  appropriate densities of states and integrating over energy.
That is, we define the quasiparticle and phonon densities as:
\begin{equation}
{\mathcal{N}}_{\mathrm{qp}} = 2\nu_F \int_{\Delta}^\infty dE\, \rho(E) f_{\mathrm{qp}}(E),
\label{eq:Nqp_def}
\end{equation}
\begin{equation}
{\mathcal{N}}_S = N_{{S}} \int_{2\Delta}^\infty d\Omega\, F_S(\Omega) n_S(\Omega),
\label{eq:NS_def}
\end{equation}
\begin{equation}
{\mathcal{N}}_s = N_{{S}} \int_{2\Delta}^\infty d\Omega\, F_{S}(\Omega) n_s(\Omega),
\label{eq:Ns_def}
\end{equation}
where $\nu_F$ is the electronic density of states at the Fermi level (including spin degeneracy), $\rho$ is the (normalized) BCS density of states, $f_{\mathrm{qp}}$ is the quasiparticle distribution function, $n_S$ is the phonon distribution function in the superconductor, $N_{{S}}$ is its atomic number density, and $F_{S}(\Omega)$ is the phonon density of states in the superconductor normalized such that $\int_0^\infty F_S(\Omega)d\Omega = 3$ (note that for simplicity we multiplied the distribution of phonons in the substrate by the phonon density of states in the superconductor). 
Then the RT equations take the form:
\begin{align}
\frac{d{\mathcal{N}}_{qp}}{dt} &= -2R {\mathcal{N}}_{qp}^2 - s {\mathcal{N}}_{qp} + \frac{2}{\tau_0^\mathrm{PB}} {\mathcal{N}}_{S},\\
\frac{d{\mathcal{N}}_{S}}{dt} &= R {\mathcal{N}}_{qp}^2 - \frac{1}{\tau_0^\mathrm{PB}} {\mathcal{N}}_{S} - \frac{{\mathcal{N}}_{S}-{\mathcal{N}}_s}{\tau_{Ss}},\\
\frac{d{\mathcal{N}}_s}{dt} &= {\mathcal{N}}_{s0} \delta(t) - \frac{{\mathcal{N}}_s-{\mathcal{N}}_S}{\tau_{sS}} - \frac{{\mathcal{N}}_s}{\tau_\mathrm{es}}, \label{eq:Nsub}
\end{align}
where the recombination coefficient $R$ and the pair-breaking lifetime of phonons in the superconductor are related to $\tau_\mathrm{el}$ as
\begin{equation}\label{eq:RtPB}
    R= \frac{1}{\tau_\mathrm{el}}\frac{2\Delta^2}{\nu_F \Omega_D^3}, \qquad \frac{1}{\tau_0^\mathrm{PB}}= \frac{1}{\tau_\mathrm{el}} \frac{\pi \Delta \nu_F}{N \beta \Omega_D^3} = \frac{1}{\tau_\mathrm{ph}}\frac{\pi \Delta}{\Omega_D}
\end{equation}
with $\beta= b/\overline{\alpha}^2$ (cf. Sec.~\ref{sec:E-PH}). The parameter $s$ phenomenologically accounts for quasiparticle trapping (and/or recombination of excess quasiparticles with background ones). {The term proportional to $\delta(t)$ in Eq.~\eqref{eq:Nsub} represents a sudden generation of substrate phonons at $t=0$.}

We are interested in the dynamics at long time after a pulse injection of phonons in the substrate. Due to the finite escape time $\tau_\mathrm{es}$, most phonons will have left the system, so we expect all the densities to be small. Then we can neglect the terms proportional to $R$ under the assumptions $R{\mathcal{N}}_{qp} \ll s$ and $R \tau_{{Ss}} {\mathcal{N}}_{qp}^2 \ll {\mathcal{N}}_s$. Moreover, if the time scale characterizing the long-time dynamics is long compared to $1/s$ and $\max\{\tau_0^\mathrm{PB}, \, \tau_{Ss}\}$, in the first two equations we can neglect the time derivative terms in the left-hand sides. Solving the resulting equations for ${\mathcal{N}}_{qp}$ and ${\mathcal{N}}_S$ in terms of ${\mathcal{N}}_s$ we find
\begin{equation}
    {\mathcal{N}}_{qp} = \frac{{2}}{s \tau_0^\mathrm{PB}} {\mathcal{N}}_S, \qquad {\mathcal{N}}_S = \frac{1/\tau_{{Ss}}} {1/\tau_0^\mathrm{PB}+1/\tau_{Ss}}{\mathcal{N}}_s \, .
\end{equation}
Substituting the last expression into Eq.~\eqref{eq:Nsub} we immediately find that ${\mathcal{N}}_s$ decays exponentially over time at a rate $\gamma$ given by
\begin{equation}\label{eq:gamma}
    \gamma = \frac{1}{\tau_\mathrm{es}} + \frac{1}{\tau_{sS}}\frac{1}{1+\tau_0^\mathrm{PB}/\tau_{Ss}} \, .
\end{equation}
If the last term is small compared to the first one, the decay rate is to a good approximation determined by the escape time $\tau_\mathrm{es}$; a sufficient condition for this to be the case is $\tau_{sS} \gg \tau_\mathrm{es}$. To estimate $\tau_{sS}$, we proceed as for the estimation of $\tau_{sm}$ in Sec.~\ref{sec:mstr}, see Eq.~\eqref{eq:transmission_times} where we replace $m\to S$. Considering Al on Si with typical thickness $d_s \sim 500\,\mu$m, using the data in Tables~\ref{tab:material_properties} and \ref{tab:avg_coefficients}, the estimate would be $\sim 0.4\,\mu$s; however, given the typical surface area of an Al-only qubit, which is of the order $0.1\,$mm$^2$, in comparison to the chip area of order 100~mm$^2$, the transmission probability is reduced by at least three orders of magnitude {(in the notation of Sec.~\ref{sec:mstr}, the coverage fraction is $p=10^{-3}$)}, so our actual estimate is $\tau_{sS}\gtrsim 0.4\,$ms, a few times larger than $\tau_\mathrm{es}$ for chips glued to the sample holder, see Sec.~\ref{sec:Phonon escape time}; note that for qubits with capacitor pads made of higher-gap material, as in Refs.~\cite{iaia_etal,Modeling}, the Al region is at least a few orders of magnitude smaller and hence $\tau_{sS}$ much longer. Moreover, the factor $\tau_0^\mathrm{PB}/\tau_{Ss}$ can further suppress the $1/\tau_{sS}$ contribution to $\gamma$. Using Eq.~\eqref{eq:RtPB} {and the values for Al in Table~\ref{tab:eph_prefactors}} we find $\tau_0^\mathrm{PB} \sim {0.27}\,$ns, and taking again Al on Si with typical thickness $d_S \sim {100}\,$nm we find $\tau_{Ss} \sim {0.36}\,$ns, which in fact gives a suppression factor of about {1.75}. We hence conclude that $\gamma \simeq 1/\tau_\mathrm{es}$.

\section{Electron--phonon interaction in disordered metals}
\label{app:dirty_eph}

The quadratic form used in the main text for the Eliashberg function $\alpha^2(\Omega)F(\Omega)$ can be generalized to account for electron impurity scattering, in which case the power law is different:
\begin{equation}
    \alpha^2(\Omega)F(\Omega)=b_p\Omega^p,
    \label{eq:dirty_A2F}
\end{equation}
where $p=2$ applies to clean metals ($b_2 \equiv b$), while $p=1$ and $p=3$ describe disordered metal with fixed scatterers and impurities that move with the lattice, respectively -- see Ref.~\cite{basko.2019} and references there. With this notation, the electron kernel in  Eq.~\eqref{eq:Iel} becomes
\begin{equation}
    \frac{\Omega^p}{\tau_{\mathrm{el},p}\Omega_D^{p+1}},
    \qquad
    \tau_{\mathrm{el},p}
    =
    \frac{\hbar}{2\pi b_p\Omega_D^{p+1}}.
    \label{eq:dirty_el_kernel}
\end{equation}
Keeping $F(\Omega)=\beta\Omega^2$ for the phonon density of states, the phonon kernel in Eq.~\eqref{eq:Iph} is modified as
\begin{equation}
    \frac{1}{\tau_{\mathrm{ph},p}\Omega_D}
    \left(\frac{\Omega}{\Omega_D}\right)^{p-2},
    \qquad
    \tau_{\mathrm{ph},p}
    =
    \frac{\hbar N\beta}{2\pi\nu_F b_p\Omega_D^{p-1}}.
    \label{eq:dirty_ph_kernel}
\end{equation}

With the modifications above, repeating the considerations of Sec.~\ref{sec:analytic} lead to the same expression for $n(\Omega)$ in Eqs.~\eqref{eq:nlO} and \eqref{eq:n1} upon the substitution $\tilde{\Omega} \to \tilde{\Omega}_p(\Omega)$, 
with
\begin{equation}
    \tilde\Omega_p(\Omega)
    =
    \frac{\tau_{\mathrm{ph},p}\Omega_D}{\bar\tau_{ms}}
    \left(\frac{\Omega_D}{\Omega}\right)^{p-2}.
    \label{eq:dirty_tildeOmega_p}
\end{equation}
Calculation of the down-conversion efficiency $R_{qp}$ could then proceed as in Sec.~\ref{sect:DW_eff}. However, quantitative estimates would require the knowledge of the prefactors $b_1$ or $b_3$; while these could be in principle determined experimentally (for instance by the power dissipation measurements mentioned in Sec.~\ref{sec:E-PH}), we leave detailed exploration of the disordered metal regime to future work.

\section{Numerical method}\label{app:numerical}

We numerically solve the coupled nonlinear integral equations
\eqref{eq:f_st} and \eqref{eq:k_eq_eff}
using a Newton-Raphson method. The implementation follows the
approach of Ref.~\cite{Paul}, adapted to normal metals.
The energy variables for both electron and phonon distributions
are discretized on the interval $[0,\Lambda]$, where the cutoff {
$ \Lambda=\min\!\left[
 \max(10\Omega_{\rm inj},20\tilde E),\Omega_D
 \right]$},
with $\tilde E$ as defined in Eq.~\eqref{eq:tilde_E}.
We employ $n=6001$
{grid points with uniform intervals}.
The same grid is used for the phonon energy $\Omega$.

The monochromatic phonon injection
$n_{\rm inj}(\Omega)=
\tilde n_{\rm inj}\delta(\Omega-\Omega_{\rm inj})$,
{with
$\tilde n_{\rm inj}=\tau_{\rm es}\tilde g$,}
is discretized by distributing the injection over the
two grid points enclosing $\Omega_{\rm inj}$.
If $\Omega_k<\Omega_{\rm inj}<\Omega_{k+1}$, we set
{
\begin{align}
 N_{{\rm inj},k}
 &=\frac{\tilde n_{\rm inj}(1-\alpha)}{w_{\Omega,k}},\\
 N_{{\rm inj},k+1}
 &=\frac{\tilde n_{\rm inj}\alpha}{w_{\Omega,k+1}},
\end{align}
}
where
\[
 \alpha=
 \frac{\Omega_{\rm inj}-\Omega_k}
 {\Omega_{k+1}-\Omega_k},
\]
and $w_{\Omega,i}$ are the trapezoidal integration weights:
\begin{equation}
 w_i=
 \begin{cases}
 (E_1-E_0)/2, & i=0,\\
 (E_{i+1}-E_{i-1})/2, & 0<i<n,\\
 (E_n-E_{n-1})/2, & i=n.
 \end{cases}
\end{equation}
This construction preserves the normalization and first moment
of the delta function:
{
\begin{equation}
 \sum_j w_{\Omega,j}\Omega_j^qN_{{\rm inj},j}
 =\tilde n_{\rm inj}\Omega_{\rm inj}^q,
 \qquad q=0,1.
\end{equation}
}

We collect 
the values of the distribution functions at the discrete energy points
into the vector $\mathbf{x} = (f_0, \ldots, f_n, n_0, \ldots, n_n)^T \in \mathbb{R}^{2(n+1)}$. The system of coupled integral equations becomes $\mathbf{F}(\mathbf{x}) = 0$, where
\begin{align}
F_m(\mathbf{x}) &= I_{el}[m; \mathbf{f}, \mathbf{n}], & m = 0, \ldots, n, \\
F_{n+1+m}(\mathbf{x}) &= I_{ph}[m; \mathbf{f}, \mathbf{n}]\! - \! \frac{n_m - N_{\text{inj},m}}{\bar{\tau}_{ms}}, & m = 0, \ldots, n.
\end{align}
We solve this using the damped Newton-Raphson algorithm:
\begin{equation}
\mathbf{x}^{(k+1)} = \mathbf{x}^{(k)} - \alpha\, \mathbf{J}^{-1}(\mathbf{x}^{(k)})\, \mathbf{F}(\mathbf{x}^{(k)}),
\end{equation}
with damping factor $\alpha = 0.9$. The Jacobian matrix $J_{mn} = \partial F_m/\partial x_n$ is computed analytically from the discretized collision integrals to ensure numerical stability and faster convergence. {We stop when the norm of the residual is below
$5\times10^{-7}$}.

Since the Newton-Raphson method can be sensitive to the initial guess,
we verified the robustness of our numerical framework by implementing
two different initialization strategies.

In the first approach, we initialize the distributions at each parameter
point $(\tilde{g}, \Omega_\text{inj})$ using discretized versions of our
analytical approximations: the Fermi-like distribution from
Eq.~\eqref{eq:approx_distributions} for $f(E)$, and the Bose-like function
with escape corrections from Eq.~\eqref{eq:n_approx} for $n(\Omega)$.
With this initial guess, the algorithm typically converges in
{2--3 Newton steps. We use this initialization
independently at each parameter point, without warm starts.}

The second scheme relies on a Picard iteration method derived from the
electron kinetic equation in the
{fast phonon escape} limit
($\bar{\tau}_{ms} \to 0$).
By isolating $f(E)$ in the discretized version of
Eq.~\eqref{eq:schematic_electron} for the energy range
$E < \Omega_\text{inj}$,
{and neglecting contributions from electronic
energies above $\Omega_\text{inj}$ in this initialization,}
we obtain the mapping:
\begin{equation}
f_m^{(k+1)} = \frac{\bar{E}^3 + \int_{E_m}^{\Omega_\text{inj}} (\Omega - E_m)^2\, f_\Omega^{(k)}\, d\Omega - \bar{E}^3\, f_{\Omega_\text{inj} - E_m}^{(k)}}
{2\left[\bar{E}^3 + E_m^3/6 + \int_0^{\Omega_\text{inj}} \Omega^2\, f_\Omega^{(k)}\, d\Omega\right]},
\end{equation}
starting from $f_m^{(0)} = 1/(2 + E_m)$. The term $f_{\Omega_\text{inj} - E_m}^{(k)}$ is evaluated at the grid point closest to the exact energy difference. Once the Picard iteration yields a converged electron distribution $f$, we initialize the corresponding phonon distribution using the first-order expansion:
\begin{equation}
n(\Omega) \simeq n_0(\Omega) + \bar{\tau}_{ms} I_\mathrm{ph} \left\{n_0, f\right\},
\end{equation}
where $n_0(\Omega) = \tilde{n}_{\rm inj}\,\delta(\Omega-\Omega_{\rm inj})$.

{We use the second initialization as an independent
check at ten representative parameter points for Cu, applying the
full Picard procedure separately at each point. For this comparison,
both initialization schemes are followed by Newton iterations with
damping $\alpha=0.9$.}
For {these ten} representative points, solutions
obtained from both initialization schemes agree within
{$10^{-6}$} %{$10^{-4}\%$}, 
based on the normalized $L^2$ norm of their difference
{evaluated separately for $f$ and $n$}.

\section{Numerical distribution functions}
\label{app:more_sol}

\begin{figure}[htbp]
    \centering
    \includegraphics[width=0.99\columnwidth]{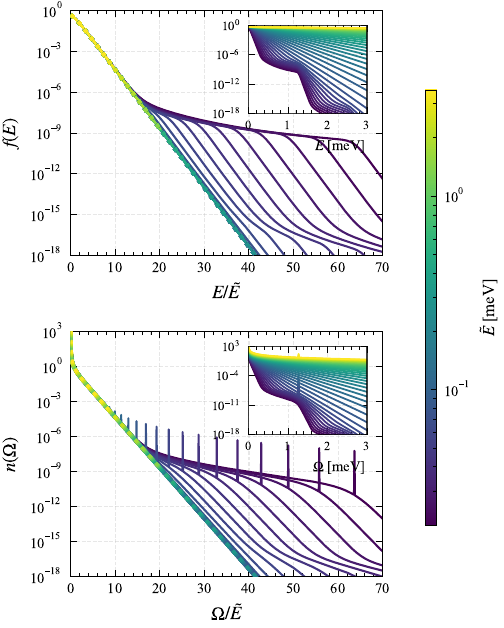}
    \caption{
    Electron distribution $f(E)$ {(top panel)} and phonon distribution $n(\Omega)$ {(bottom panel)} in a Cu film under steady, monochromatic phonon injection at $\Omega_{\rm inj} = {1.265} \, \mathrm{meV}$.
    The dashed lines correspond to the electron distribution in Eq.~\eqref{eq:approx_distributions} in {the top} panel and to the phonon distribution in Eq.~\eqref{eq:n_approx} in {in the bottom one}.
    }
    \label{fig:Cu_distributions2}
\end{figure}

We present here additional examples of numerical results for the distribution function to complement Fig.~\ref{fig:Cu_distributions}. For a higher injection energy, $\Omega_\mathrm{inj} = {1.265}\,$meV,  the results are in fact similar, see Fig.~\ref{fig:Cu_distributions2}, except for the slower-than-exponential decrease extending to higher energies because of the higher $\Omega_\mathrm{inj}$. Additionally, we check the robustness of our findings to the spectrum of the injected {phonons}. As an example, we consider a Gaussian spectrum, which reduces to a $\delta$-function in the limit of vanishing standard deviation, $\sigma\to 0$:
\begin{equation}\label{eq:Gauss_inj}
    n_{\rm inj}(\Omega)
    =
    A_\sigma
    \exp\!\left[-\frac{(\Omega-\Omega_{\rm inj})^2}{2\sigma^2}\right]\theta(\Omega)
\end{equation}
with $\theta$ the step function. To compare the results for different injection spectra, for a given $\sigma$ we fix the amplitude $A_\sigma$ so that the resulting energy scales $\tilde{E}$, as defined by Eq.~\eqref{eq:Etilde_cond}, are the same,
\begin{align}
    A_\sigma = & \frac{(\pi \tilde{E})^4}{15}\Bigg\{\sqrt{\frac{\pi}{2}}\sigma \left(\Omega_\mathrm{inj}^3+3\Omega_\mathrm{inj}\sigma^2\right)\left[1+\mathrm{Erf}\left(\frac{\Omega_\mathrm{inj}}{\sqrt{2}\sigma}\right)\right]  \nonumber \\ &
    +\sigma^2 (\Omega^2_\mathrm{inj}+2\sigma^2)\exp\left(- \frac{\Omega^2_\mathrm{inj}}{2\sigma^2} \right)\Bigg\}^{-1}.
\end{align}

\begin{figure}[htbp]
    \centering
    \includegraphics[width=0.99\columnwidth]{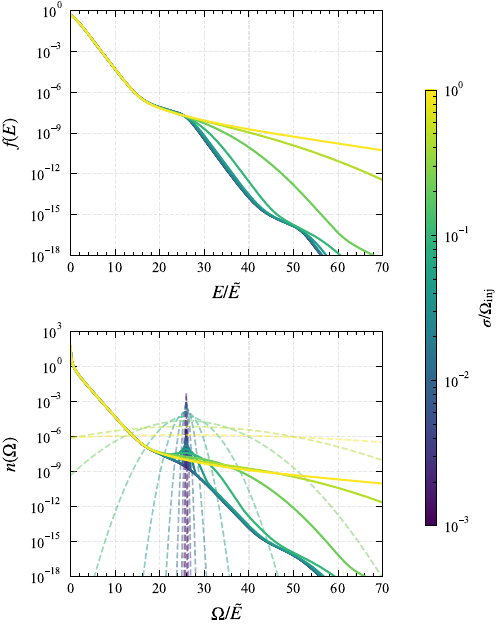}
    \caption{
    Steady-state electron (top) and phonon (bottom) distributions in a Cu film
    under phonon injection centered at $\Omega_{\rm inj} = 0.495$ meV.
    The injection rate is fixed,
    $\tilde g = {0.1425}\,{\rm meV/s}$, corresponding to
    $\tilde E \simeq {0.019} \,{\rm meV}$.
    Solid curves denote the numerical solutions, and dash-dotted curves the
    equilibrium-like approximations.
    Dashed curves in the bottom panel show the Gaussian injection profiles
    $n_{\rm inj}(\Omega)$ of Eq.~\eqref{eq:Gauss_inj} for different
    $\sigma/\Omega_{\rm inj}$, normalized to give the same $\tilde E$.
    }
    \label{fig:delta_gaussian_compare}
\end{figure}

\begin{figure}[htbp]
    \centering
    \includegraphics[width=0.99\columnwidth]{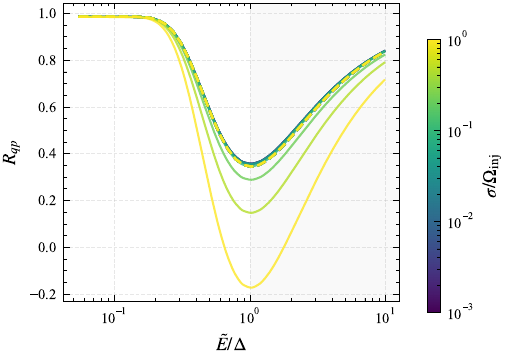}
   \caption{ 
    Down-conversion efficiency $R_{\rm qp}$,
    Eq.~\eqref{eq:Rqp_definition}, vs. $\tilde E/\Delta$ for a
    $100\,{\rm nm}$ Al film
    ($\Delta=0.18\,{\rm meV}$, $\lambda\simeq {0.42}$).
    The metal distributions are obtained numerically {(cf. Fig.~\ref{fig:delta_gaussian_compare})} for a
    Gaussian phonon injection spectrum, Eq.~\eqref{eq:Gauss_inj}, centered at
    $\Omega_{\rm inj}\simeq 0.5\,{\rm meV}$. 
    As in Fig.~\ref{fig:Rqp_vs_delta}, the shaded region at $\tilde{E}/\Delta {>1}$ is outside the validity regime of our approximation for the phonon distribution in the superconductor.
    }
    \label{fig:R_qp_sigma}
\end{figure}

As evident in Fig.~\ref{fig:delta_gaussian_compare}, even for broad injection spectra, $\sigma \lesssim \Omega_\mathrm{inj}$ the low-energy parts of the distributions are indistinguishable from those obtained for narrow injection. However, this does not imply that the down-conversion efficiency is unaffected; we plot the latter in  Fig.~\ref{fig:R_qp_sigma} as a function of $\tilde{E}/\Delta$ [cf. Fig.~\ref{fig:Rqp_vs_delta}] for various values of the broadening $\sigma$. In the region $\tilde{E}/\Delta \gtrsim 0.2$ where the efficiency is suppressed below the `perfect absorption' value, the suppression can become stronger, but only for sufficiently large broadening $\sigma/\Omega_\mathrm{inj} > 0.1$.

\section{Deviation from equilibrium-like distributions}
\label{app:steps}

We explore here briefly the high-energy structure of the electron distribution function revealed by numerics, see Sec.~\ref{sec:numerical_results} and Appendix~\ref{app:more_sol}. To that end, we adopt an iterative strategy, starting from the equilibrium-like expressions for the distributions in Eq.~\eqref{eq:approx_distributions}, which we denote here with $f^{(0)}$ and $n^{(0)}$. They are solutions to Eq.~\eqref{eq:f_st}, but not to Eq.~\eqref{eq:k_eq_eff}; using $f^{(0)}$ in the latter, we find the next iteration for $n$, which we denote with $n^{(1)}$; it has the form given in Eq.~\eqref{eq:nlO} with $n_1$ of Eq.~\eqref{eq:n_approx}. 

To find the next iteration for $f$, we write it in the form $f^{(1)} = f^{(0)}+\delta f$; we substitute this definition and $n^{(1)}$ into Eq.~\eqref{eq:f_st} to derive an approximate equation for $\delta f$. The approximations to be used originate from the observation that products of the form $f^{(0)}(1-f^{(0)}){n^{(1)}}$ can be written identically as $f^{(0)}(1-f^{(0)})(n^{(0)}+\delta n)$ with $\delta n =n^{(1)}-n^{(0)}$; only terms containing the difference $\delta n$ contribute to the equation for $\delta f$ as source terms, and the difference vanishes in the limit $\tilde{\Omega}\to 0$. This suggest 
keeping only terms linear in $\delta f$, and dropping products of the form $\delta f \delta n$; the validity of the approximations can be checked \textit{a posteriori}. The resulting equation is
\begin{equation}
    \mathcal{I}_{out}\{f^{(0)},n^{(0)}\}\delta f(E)-\mathcal{I}_{in}\{\delta f; f^{(0)}, n^{(0)} \} = \mathcal{S}\{f^{(0)},\delta n\}.
\end{equation}
We discuss next the three terms in this equation, starting with the right-hand side.

The source term can be written compactly as
\begin{align}
    \mathcal{S} \{f^{(0)},\delta n\} & = \int_0 \frac{\Omega^2 d\Omega}{{\tilde{E}^3}} \, \delta n(\Omega) \\ & \left[f^{(0)}(E+\Omega) + f^{(0)}(E-\Omega) -2f^{(0)}(E) \right]. \nonumber 
\end{align}
Hereinafter we assume all energy scales to be small compared to the Debye energy, so that the upper integration limit can be omitted. Given the structure of $\delta n$, we can naturally split the source term into two contributions, one due to the (renormalized) phonon injection and one to the equilibrium-like component,
\begin{align}
    \mathcal{S} & = \mathcal{S}_\mathrm{inj} - \mathcal{S}_{eq} \, , \label{eq:source}\\
    \mathcal{S}_\mathrm{inj} & =\frac{\pi^4}{15} \frac{\tilde\Omega}{\tilde{E}}\left(\frac{\tilde{E}}{\Omega_\mathrm{inj}}\right)^2 \tanh\left(\frac{E}{2\tilde{E}}\right) J\left(\frac{E}{\tilde{E}},\frac{\Omega_\mathrm{inj}}{\tilde{E}}\right), \label{eq:Sinj} \\
     \mathcal{S}_{eq} & = \frac{\tilde\Omega}{\tilde{E}}  \tanh\left(\frac{E}{2\tilde{E}}\right) \int_0 y dy \frac{J(E/\tilde{E},y)}{e^{y}-1} \\
     & = \frac{\tilde\Omega}{\tilde{E}}  \frac{\tanh\left(E/2\tilde{E}\right)}{\sinh(E/\tilde{E})} \bigg\{2\cosh^2\left(\frac{E}{2\tilde{E}}\right)\mathrm{Li}_2\left(-e^{-E/\tilde{E}}\right) \nonumber \\ & +\left[\frac{\pi^2}{12}+\frac{1}{4}\left(\frac{E}{\tilde{E}}\right)^2\right]\left(1+e^{-E/\tilde{E}}\right)\bigg\}, \nonumber
\end{align}
where we assumed $\tilde\Omega \ll \Omega_\mathrm{inj},\, \tilde{E}$, $\mathrm{Li}_s$ denotes the polylogarithm of order $s$, and introduced
\begin{equation}
    J(x,y) = \frac{\cosh y -1}{\cosh y + \cosh x}.
\end{equation}

The outgoing collision integral prefactor can be written as
\begin{align}\label{eq:Iout}
     \mathcal{I}_{out} = & \int_0 
\frac{\Omega^2 d\Omega}{\tilde E^3} \\ &
\left[f^{(0)}(E+\Omega) + f^{(0)}(\Omega-E) + 2n^{(0)}(\Omega)
\right] \nonumber \\
= & \, 4 \zeta(3)-2 \mathrm{Li}_3\left(-e^{-E/\tilde{E}}\right)-2 \mathrm{Li}_3\left(-e^{E/\tilde{E}}\right). \nonumber
\end{align}
Finally, the ingoing collision integral is
\begin{align}\label{eq:iin}
     \mathcal{I}_{in} & = \int_0 \frac{\Omega^2 d\Omega}{\tilde E^3}
\bigg\{
\left[ 1-f^{(0)}(E)+n^{(0)}(\Omega) \right]
\delta f(E+\Omega)
\nonumber \\
& +
\left[
f^{(0)}(E)+n^{(0)}(\Omega)
\right]
\delta f(E-\Omega)
\bigg\},
\end{align}
where for $E<0$ we define $\delta f(E) = - \delta f(-E)$.

\begin{figure}[tb]
    \centering
    \includegraphics[width=0.99\columnwidth]{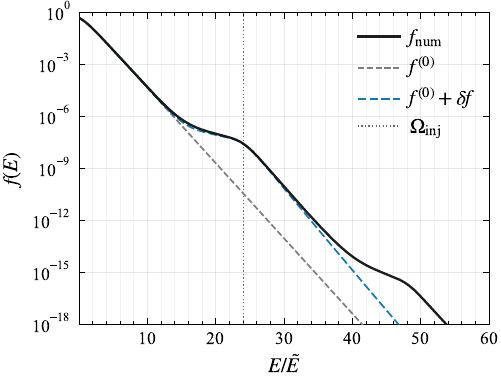}
    \includegraphics[width=0.99\columnwidth]{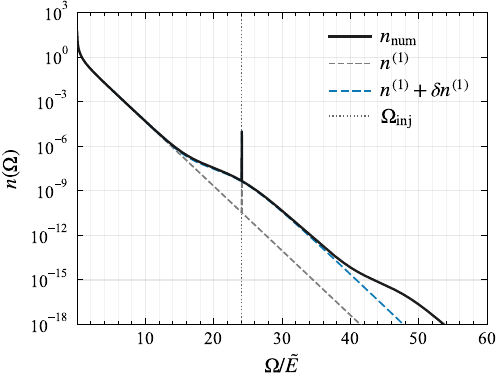}
    \caption{
    Electron and phonon distributions in Cu for 
    {$\Omega_{\rm inj} = 0.37\,\mathrm{meV}$} and
    $\tilde{g} = {0.143}\,{\rm meV}/\mathrm{s}$, corresponding to $\tilde{E} {\simeq 0.015}\,$meV.
    Top panel: electron distribution. The solid black line represents the numerical
    solution, the {short-}dashed {gray} line the equilibrium-like approximation in
    Eq.~\eqref{eq:approx_distributions}, and the {long-}dashed blue line the same approximation
    {plus} the correction of Eq.~\eqref{eq:df_he}.
    Bottom panel: phonon distribution in the same configuration. The solid black line
    represents the numerical solution, the {short-}dashed {gray} line the equilibrium-like approximation with escape correction, {Eq.~\eqref{eq:n_approx}}, and the {long-}dashed blue line the same approximation {plus} the first-order correction $\delta n$ in Eq.~\eqref{eq:dn_he}.
    }
    \label{fig:Cu_distributions_corr}
\end{figure}

From the structure of the source term it is clear that $\delta f \propto \tilde{\Omega}/\tilde{E}$ (the latter being the small parameter justifying a perturbative approach) and that there is a one-parameter family of solutions for $\delta f$ parametrized by $\Omega_\mathrm{inj}/\tilde{E}$ [cf. Eq.~\eqref{eq:Sinj}]. Here we focus on the regime $\Omega_\mathrm{inj}/\tilde{E}\gg 1$ in which the deviation from exponential decay at $E/\tilde{E}\gg 1$ is evident in the numerical results. When these two quantities are large, we can approximate the two contributions to the source term as
\begin{align}
    \mathcal{S}_\mathrm{inj} & \simeq \frac{\pi^4}{15} \frac{\tilde\Omega}{\tilde{E}}\left(\frac{\tilde{E}}{\Omega_\mathrm{inj}}\right)^2 \frac{1}{e^{(E-\Omega_\mathrm{inj})/\tilde{E}}+1},\\
    \mathcal{S}_{eq} & \simeq \frac{1}{2}  \frac{\tilde\Omega}{\tilde{E}}\left(\frac{E}{\tilde{E}}\right)^2 e^{-E/\tilde{E}}.
\end{align}
From these expressions we deduce that the injection term is dominant in a wide range of energies around $\Omega_\mathrm{inj}$; more precisely, $\mathcal{S}_\mathrm{inj} \gg \mathcal{S}_{eq}$ for $2\log(\Omega_\mathrm{inj}/\tilde{E})\ll E/\tilde{E} \ll (2\pi^4/15) e^{\Omega_\mathrm{inj}/\tilde{E}}/(\Omega_\mathrm{inj}/\tilde{E})^2$. Assuming that the ingoing collision integral can be ignored, and using that in this regime $\mathcal{I}_{out} \simeq (E/\tilde{E})^3/3$, we get
\begin{equation}\label{eq:df_he}
    \delta f(E) \simeq \frac{\pi^4}{5}\frac{\tilde\Omega}{\tilde{E}}\left(\frac{\tilde{E}}{\Omega_\mathrm{inj}}\right)^2 \left(\frac{\tilde{E}}{E}\right)^3 \frac{1}{e^{(E-\Omega_\mathrm{inj})/\tilde{E}}+1}.
\end{equation}
The correction is clearly small in the energy range where the derivation applies, $\delta f \ll 1$, justifying linearization. Nonetheless, the correction can be larger than the exponential tail of $f_0$; considering energies of order $\Omega_\mathrm{inj}$, the requirement $\delta f \gtrsim f_0$ results in the condition \(\Omega_\mathrm{inj}/\tilde{E} > -5 W_{-1}[-(\pi^4\tilde{\Omega}/10\tilde{E})^{1/5}/5]\), with $W$ the Lambert W (product logarithm) function. For the parameters in Figs.~\ref{fig:Cu_distributions} and \ref{fig:Cu_distributions2}, this expression puts the crossover from thermal-like distribution at all energies to deviation in the tails at $\Omega_\mathrm{inj}/\tilde{E} \sim 15$-$18$, consistent with the numerical results. In the top panel of Fig.~\ref{fig:Cu_distributions_corr} we compare the numerical result for $f$ to the analytical approximation $f^{(0)} + \delta f$, finding good agreement between them.

The correction to the phonon distribution $\delta n^{(1)}$ [not to be confused with the difference $\delta n$] can be obtained by keeping terms linear in $\delta f$ in Eq.~\eqref{eq:n1}, resulting in
\begin{align}
    \delta n^{(1)}(\Omega) = & \ {\frac{2}{\Omega+\tilde\Omega}}\int_0 dE\,\delta f(E) \\
& \times \left[
f^{(0)}(\Omega-E)-f^{(0)}(E+\Omega)
\right]. \nonumber
\end{align}
Focusing on the regime $\Omega \gg \tilde{E}$, we approximate the term in square brackets as $1/[e^{(\Omega-E)/\tilde{E}}+1]$ and $\delta f$ as in Eq.~\eqref{eq:df_he}. Then we consider separately the cases $\Omega < \Omega_\mathrm{inj}$ and $\Omega > \Omega_\mathrm{inj}$ (with $|\Omega-\Omega_\mathrm{inj}| \gg \tilde{E}$); we find that the leading contributions can be captured by the interpolating formula
\begin{align}\label{eq:dn_he}
    \delta n^{(1)}(\Omega) & \simeq \frac{\pi^4}{5} \frac{\tilde\Omega}{\tilde{E}} \left(\frac{\tilde{E}}{\Omega_\mathrm{inj}}\right)^2\frac{\tilde{E}}{\Omega} \\
    & \left[\left(\frac{\tilde{E}}{\Omega_\mathrm{inj}}\right)^2-\left(\frac{\tilde{E}}{\Omega}\right)^2\right]\frac{1}{e^{(\Omega-\Omega_\mathrm{inj})/\tilde{E}}-1}. \nonumber
\end{align}
The bottom panel in Fig.~\ref{fig:Cu_distributions_corr} shows that the approximation $n^{(1)} + \delta n^{(1)}$ quantitatively captures the behavior of the numerical result for $n$ up to the first `step'.

We note that as long as neglecting the ingoing collision integral is a good approximation, using Eqs.~\eqref{eq:source}-\eqref{eq:Iout} a closed form for the correction is simply given by $\delta f \simeq \mathcal{S}/\mathcal{I}_{out}$. We do not investigate here in detail the validity of the approximation; we remark though that at sufficiently high energy the approximation fails. In fact, one can show that for $E>\Omega_\mathrm{inj}\gg \tilde{E}$ the main contribution to $\mathcal{I}_{in}$ originates from the last term in Eq.~\eqref{eq:iin},
\begin{align}
    \mathcal{I}_{in} & \simeq \int_0^{E} \frac{\Omega^2 d\Omega}{\tilde E^3} n^{(0)}(\Omega) \delta f(E-\Omega) \\
    & \simeq \frac{\pi^4}{5} \frac{\tilde{\Omega}}{\tilde{E}} \left(\frac{\tilde{E}}{\Omega_\mathrm{inj}}\right)^2 e^{-(E-\Omega_\mathrm{inj})/\tilde{E}}\int_0^{E-\Omega_\mathrm{inj}} \!\!d\Omega \frac{\Omega^2}{(E-\Omega)^2} \nonumber \\
    & = \frac{\pi^4}{5} \frac{\tilde{\Omega}}{\tilde{E}} \left(\frac{\tilde{E}}{\Omega_\mathrm{inj}}\right)^2 e^{-(E-\Omega_\mathrm{inj})/\tilde{E}} \left[\log\frac{E}{\Omega_\mathrm{inj}} \right. \nonumber \\ & \qquad \qquad \qquad \left. +\frac{(E-3\Omega_\mathrm{inj})(E-\Omega_\mathrm{inj})}{2\Omega_\mathrm{inj}^2}\right], \nonumber
\end{align}
where in going from the first to the second line we have {used Eq.~\eqref{eq:df_he} and} approximated $n^{(0)}/(e^{(E{-\Omega}-\Omega_\mathrm{inj})/\tilde{E}}+1) \simeq e^{-(E-\Omega_\mathrm{inj})/\tilde{E}}\theta(E-\Omega_\mathrm{inj}-\Omega)$. Using this result, we find that $\mathcal{I}_{out}\delta f \simeq \mathcal{I}_{in}$ at $E \simeq 2.2 \Omega_\mathrm{inj}$. This shows that the higher-energy structure of the tails cannot be captured when dropping $\mathcal{I}_{in}$.

\section{Relation between substrate phonons and superconducting device phonons}
\label{app:ns_nS_relation}

In this Appendix we derive the relation between the phonon distribution in the substrate, \(n_s(\Omega)\), and the phonon distribution inside the superconducting device, \(n_S(\Omega)\). from the coupled kinetic equations for the substrate, the normal metal, and the superconducting film.
Taking into account the latter, in the steady-state Eq.~\eqref{eq:ns_st} for the substrate phonon distribution $n_s$ is replaced by
\begin{equation}
g(\Omega)
-\frac{n_s(\Omega)}{\tau_{\mathrm{es}}}
-\frac{n_s(\Omega)-{n(\Omega)}}{\tau_{sm}}
-\frac{n_s(\Omega)-{n_S(\Omega)}}{\tau_{sS}}
=0.
\end{equation}
The last {term} 
on the left-hand side account for superconductor phonons ($n_S$) entering the substrate
and substrate phonons leaving towards the superconductor at rate $1/\tau_{sS}$. For the phonons in the superconductor, the kinetic equation analogous to Eq.~\eqref{eq:n_st} reads
\begin{equation}\label{eq:stationary_equations_superconductor_app}
I_{\mathrm{pq}}\{n_S,f_\mathrm{q}\}
-\frac{n_S(\Omega)-{n_s(\Omega)}}{\tau_{Ss}}=0
\end{equation}
with \(I_{\mathrm{pq}}\) accounting for phonon--quasiparticle interaction in the superconducting film. These equations should be supplemented by that for the quasiparticle distribution function $f_\mathrm{q}$, $I_\mathrm{qp}\{n_S, f_\mathrm{q}\} =0$, analogous to Eq.~\eqref{eq:f_st}.

Expressions for the two collision integrals $I_\mathrm{pq}$ and $I_\mathrm{qp}$ can be found for instance in Refs.~\cite{kaplan1976,Paul}; they include quasiparticle generation by phonons with energy above the pair-breaking threshold $2\Delta$, recombination with emission of a phonon above the same energy, and quasiparticle--phonon scattering. At low quasiparticle density, recombination is slow compared to the other two processes~\cite{kaplan1976,Paul} (cf. Appendix~\ref{app:escape}), so we ignore it. 
We can also ignore scattering so long as the quasiparticles are cold, that is, their typical energy is at most of order $\Delta$; this condition restricts our considerations to $\tilde{E} \lesssim \Delta$. Under this assumptions we approximate $I_\mathrm{pq}$ as
\begin{equation}
I_{\mathrm{pq}}\{n_S,f_\mathrm{q}\}
\simeq
-\frac{1}{\tau_{\mathrm{pb}}(\Omega)} n_S(\Omega),
\label{eq:Icoll_pb_app}
\end{equation}
where \(\tau_{\mathrm{pb}}(\Omega)\) is the pair-breaking time for a phonon
of energy \(\Omega>2\Delta\) defined in Eq.~\eqref{eq:pb_rate_def_main} in terms of the spectral density~\cite{LectNotes}
\begin{equation}
S_+(x)
=
(x+2)\,E\!\left(\frac{x-2}{x+2}\right)
-
\frac{4x}{x+2}\,
K\!\left(\frac{x-2}{x+2}\right)
\label{eq:Splus}
\end{equation}
for \(x>2\) and \(S_+(x)=0\) otherwise, with \(E\) and \(K\) the complete elliptic integrals of the second and first kind, respectively. {For later use, we mention here the approximations \(S_+(x) \simeq \pi[1+(x-2)/4]\) for $x-2 \ll 2$ and \(S_+(x) \simeq x\) for \(x\gg 2\).}
Substituting Eq.~\eqref{eq:Icoll_pb_app} into 
{Eq.~\eqref{eq:stationary_equations_superconductor_app}} we obtain Eq.~\eqref{eq:nS_of_ns_general_main}.

\section{Approximate formula for the down-conversion efficiency}
\label{app:Rqp_minimum}

{We give here some details on the derivation of Eq.~\eqref{eq:Rlambda} for \(\mathcal{R}_\lambda\). Using Eq.~\eqref{eq:approx_distributions} in Eq.~\eqref{eq:Rqp_n}, which is accurate up to corrections small in \(\tilde\Omega/\Omega_\mathrm{inj}\) and \(\tilde\Omega/\tilde E\), we get Eq.~\eqref{eq:RqpRlambda} with}
\begin{equation}
    \mathcal{R}_\lambda(x) \equiv A_\lambda \frac{1+\pi\lambda}
{\pi\lambda} \frac1x \mathcal{S}_\lambda(x),
\end{equation}
where {$x=\tilde{E}/\Delta$ and}
\begin{equation}
   \mathcal{S}_\lambda(x) = \int_{2/x}^{\infty}\!dy\,
\frac{y^2}{e^y-1} \frac{\lambda S_+(xy)} {1+\lambda S_+(xy)}
\end{equation}
{with $y=\Omega/\tilde{E}$.}
This integral can be evaluated approximately in the limits \(x\to 0\) and \(x\to\infty\), as we now discuss.

{For \(x\to 0\), the exponential in the numerator is always large compared to unity and ensures that the main contribution to the integral comes for the region near the lower integration limit; with the change of integration variable \(y= 2/x + u\) we have}
\begin{equation}
 \mathcal{S}_\lambda(x) \simeq \frac{e^{-2/x}}{x^2} \int_{0}^{\infty}\!du\, \left(2 + x u \right)^2    e^{-u} \frac{\lambda S_+ (2 + xu)}{1+\lambda S_+ (2 + xu)}.
\end{equation}
Expanding the integrand to linear order in \(x\) [cf. Appendix~\ref{app:ns_nS_relation}] and performing the integral we find
\begin{equation}\label{eq:Sl0}
 \mathcal{S}_\lambda(x) \simeq \frac{e^{-2/x}}{x^2} \frac{\pi\lambda}{1+\pi\lambda} \left(4 + 4x +\frac{1}{1+\lambda\pi}x\right).
\end{equation}
For \(x\to \infty\) the lower integration limit can be set to zero and we can approximate \(S_+(xy) \simeq xy \); expanding the integrand a next-to-leading order in \(1/x\) we find
\begin{equation}\label{eq:Sli}
 \mathcal{S}_\lambda(x) \simeq 2\zeta(3) - \frac{\pi^2}{6}\frac{1}{\lambda x}.
\end{equation}
To construct an interpolation between the two limits, we note that formally taking \(x,\,\lambda \to \infty\) in Eq.~\eqref{eq:Sl0} gives {\(4/x-3/\pi\lambda x\), whose subleading term does not match} the subleading term in Eq.~\eqref{eq:Sli};
this suggests dropping the \(\lambda\)-dependent subleading corrections in those equations and combining them as to give Eq.~\eqref{eq:Rlambda}. We have checked numerically that this is a good approximation (deviation less than 2\%) for $\lambda$ down to 0.5.

\bibliography{bib_draft}

\end{document}